\documentclass[12pt]{article}
\usepackage{amssymb}
\usepackage{graphicx}
\usepackage{amsmath,amscd}
\def\be{\begin{eqnarray}}
\def\ben{\begin{eqnarray*}}
\def\ee{\end{eqnarray}}
\def\een{\end{eqnarray*}}
\def\Tr{{\rm Tr}}
\def\Lag{\mathcal{L}}

\def\D{\mathcal{D}}

\def\=:{=\hspace{-.7em}\raisebox{1.1ex}{.}\hspace{.1em}\raisebox{-0.2ex}{.} }
\newcommand{\NF}{N_{\rm F}}
\newcommand{\NC}{N_{\rm C}}

\makeatletter
\@addtoreset{equation}{section}

\renewcommand{\thefootnote}{\fnsymbol{footnote}}
\makeatother

\makeatletter
\newcommand{\thetablename}{Table}
\def\fnum@table{\thetablename\ \thetable}
\makeatother

\begin{document}
\thispagestyle{empty}
\begin{flushright}
TIT/HEP--530 \\
{\tt hep-th/0412048} \\
December, 2004 \\
\end{flushright}
\vspace{3mm}

\begin{center}
{\Large \bf 
Instantons in the Higgs Phase 
} 
\\[12mm]
\vspace{5mm}

\normalsize
{\large 
Minoru~Eto}\!\!
\footnote{\it  e-mail address: 
meto@th.phys.titech.ac.jp
}, 
  {\large 
Youichi~Isozumi}\!\!
\footnote{\it  e-mail address: 
isozumi@th.phys.titech.ac.jp
}, 
  {\large 
Muneto~Nitta}\!\!
\footnote{\it  e-mail address: 
nitta@th.phys.titech.ac.jp
}, \\
  {\large 
 Keisuke~Ohashi}\!\!
\footnote{\it  e-mail address: 
keisuke@th.phys.titech.ac.jp
} 
~and~~  {\large 
Norisuke~Sakai}\!\!
\footnote{\it  e-mail address: 
nsakai@th.phys.titech.ac.jp
} 

\vskip 1.5em

{ \it Department of Physics, Tokyo Institute of 
Technology \\
Tokyo 152-8551, JAPAN  
 }
\vspace{15mm}

{\bf Abstract}\\[5mm]
{\parbox{13cm}{\hspace{5mm}
When instantons are put into the Higgs phase, 
vortices are attached to instantons. 
We construct such composite solitons as 
$1/4$ BPS states in five-dimensional 
supersymmetric $U(N_{\rm C})$ gauge theory 
with $\NF(\ge\NC)$ fundamental hypermultiplets. 
We solve
the hypermultiplet BPS equation 
and show that all $1/4$ BPS solutions 
are generated by an $\NC\times\NF$ matrix 
which is holomorphic in two complex variables, 
assuming the vector multiplet BPS equation 
does not give additional moduli. 
We determine the total moduli space 
formed by topological sectors patched together 
and work out the multi-instanton solution 
inside a single vortex with complete moduli. 
Small instanton singularities are interpreted as 
small sigma-model lump singularities inside the vortex. 
The relation between monopoles 
and instantons in the Higgs phase is also clarified 
as limits of calorons in the Higgs phase.
Another type of instantons stuck at an intersection of 
two vortices 
and dyonic instantons in the Higgs phase are also discussed. 

}}
\end{center}
\vfill
\newpage
\setcounter{page}{1}
\setcounter{footnote}{0}
\renewcommand{\thefootnote}{\arabic{footnote}}



\section{Introduction
\label{intro}
}

Instantons have attracted much attention 
and have been applied to a wide variety of subjects 
in physics and mathematics 
since their discovery~\cite{Belavin:1975fg}. 
The method to construct multiple instanton solutions 
was established by Atiyah, Hitchin, 
Drinfeld and Manin (ADHM)~\cite{Atiyah:1978ri} 
and it was shown that 
the moduli space of instantons is a hyper-K\"ahler 
manifold.
In supersymmetric (SUSY) gauge theories,
instantons play crucial roles 
as a tool to study non-perturbative effects.  
Instanton calculus determines 
the exact superpotential 
in the low-energy effective action 
of ${\cal N}=1$ SUSY QCD~\cite{Affleck:1983mk}.
Seiberg and Witten presented the exact effective action 
of ${\cal N}=2$ SUSY QCD whose prepotential 
contains non-perturbative terms 
as instanton corrections to all orders~\cite{SW}.

The instanton solutions that are used in these 
studies of nonperturbative effects are called 
constrained instantons, which become solutions 
of the field equation only with scale-fixing 
source 
terms~\cite{Affleck}. 
This method is motivated by the necessity to fix 
the scale for instantons in the 
presence of vacuum expectation values of the Higgs field. 
This complication arises as a result of a generalized 
version of the well-known theorem by Derrick~\cite{Derrick}, 
which states that gauge theories coupled to nontrivial scalar 
fields do not allow any finite energy solution with 
four co-dimensions~\cite{MantonSutcliffe}. 
Therefore instantons have to possess an infinite amount of 
energy as solutions of the source-free field equation, 
if the gauge fields are in the Higgs phase. 
This situation is quite similar to monopoles in the 
Higgs phase which have to accompany vortices because of the 
Meissner effect~\cite{Tong:2003pz}--\cite{Kneipp}. 
In the presence of the Fayet-Iliopoulos term for the 
$U(1)$ factor gauge group, vortices are allowed to exist. 
Then it is expected to be energetically favorable 
for instantons in the Higgs phase to accompany vortices. 
Recently it has been suggested by 
Hanany and Tong \cite{Hanany:2004ea} that 
there should be a solution 
as a composite state of an instanton and vortices, 
quite similar to the monopole in the Higgs phase. 
Such a composite state is expected to be realized as 
a lump~\cite{Ward:1985ij,Stokoe:1986ic}  
(or a sigma model 
instanton~\cite{Polyakov:1975yp,Perelomov:1987va}) 
in the effective field theory on the vortex 
world-volume~\cite{Hanany:2004ea}.

Instantons (without vortices accompanied) 
in SUSY gauge theories become 
$1/2$ Bogomol'nyi-Prasad-Sommerfield (BPS) states, 
preserving a half of SUSY,  
if it is embedded in the Euclidean four space of 
the $d=4+1$ space-time. 
In string theory these BPS instantons 
can be realized as D$p$-branes on 
D$(p+4)$-branes in type IIA/IIB string~\cite{Wi}, 
and this brane configuration gives a clear 
physical interpretation of the ADHM constraints 
as the F- and D-flatness conditions in 
the SUSY gauge theory on D$p$-brane world-volume. 
Compactification of the small instanton singularity 
in the ADHM moduli space~\cite{KN} 
was understood by Nekrasov and Schwartz~\cite{NS} 
as non-commutative instantons. 
In the brane picture this phenomenon corresponds to the 
presence of a self-dual NS-NS $B$-field background 
on the D$(p+4)$-brane world-volume. 
Moreover, a direct calculation of ${\cal N}=2$ 
Seiberg-Witten prepotential was given by Nekrasov 
using the instanton counting~\cite{Nekrasov:2002qd}.

The purpose of this paper is to discuss 
instantons attached to vortices, 
when instantons are placed in the Higgs phase 
of the five-dimensional SUSY $U(N_{\rm C})$ gauge theory 
with $N_{\rm F}(\ge N_{\rm C})$ flavors of hypermultiplets 
in the fundamental representation. 
We show that 
composite states of instantons and 
vortices\footnote{
Since vortices (instantons) are defined as
solitons with codimension two (four),
they are membranes (particles) with
2+1 (0+1) dimensional world-volume in $d=4+1$ spacetime.
}
are $1/4$ BPS states. 
We solve the hypermultiplet 
BPS equations for these states. 
Assuming the vector multiplet BPS equation has a 
unique solution without additional moduli, 
we find that solutions are completely generated 
by a holomorphic $N_{\rm C}\times N_{\rm F}$ 
matrix function of 
two complex variables made of 
four codimensions of solitons. 
We call this matrix the {\it moduli matrix}. 
We find the total moduli space containing all 
topological sectors with all possible boundary conditions.
We also find that other 1/4 BPS configurations 
containing walls, vortices and monopoles \cite{Isozumi:2004vg}
are completely included in 
the 1/4 BPS states of instantons and vortices.
The moduli matrix for multiple instantons inside 
a single vortex is specified by using the effective theory 
on the vortex. 
We also obtain calorons (periodic instantons) 
in the Higgs phase and 
clarify their relation to instantons and monopoles 
in the Higgs phase. 
We also find another type of instantons which are
stuck at an intersection of two vortices.
Dyonic instantons in the Higgs phase 
and other related issues are also discussed. 

A key point of our discussion is to 
consider a 1/2 BPS vortex as a host soliton 
for certain class of composite 1/4 BPS states. 
The effective theory on 
the world volume of a single vortex 
is the SUSY ${\bf C}P^N$ 
model~\cite{Hanany:2003hp}--\cite{Eto:2004ii}. 
An instanton in the Higgs phase is 
realized as a $1/2$ BPS lump 
in this effective theory on the vortex \cite{Hanany:2004ea} 
as stated above. 
This is similar to a recent discovery of 
a monopole in the Higgs phase (a confined monopole)
\cite{Hanany:2004ea}, \cite{Tong:2003pz}--\cite{Kneipp}
as a $1/4$ BPS state,
which turns out to be a composite state 
of a monopole attached to vortices and 
is realized as a $1/2$ BPS kink~\cite{kinks} 
in the vortex effective action~\cite{Tong:2003pz}. 
In \cite{Isozumi:2004vg} 
we have shown all the solutions 
are generated by the moduli matrix 
which is holomorphic with respect to 
a single holomorphic coordinate on the wall world volume. 
The moduli matrix 
contains all moduli parameters 
in all the different topological sectors of the solitons.

\begin{figure}[t] 
\label{diagram}
\begin{center}
\includegraphics[width=15cm]{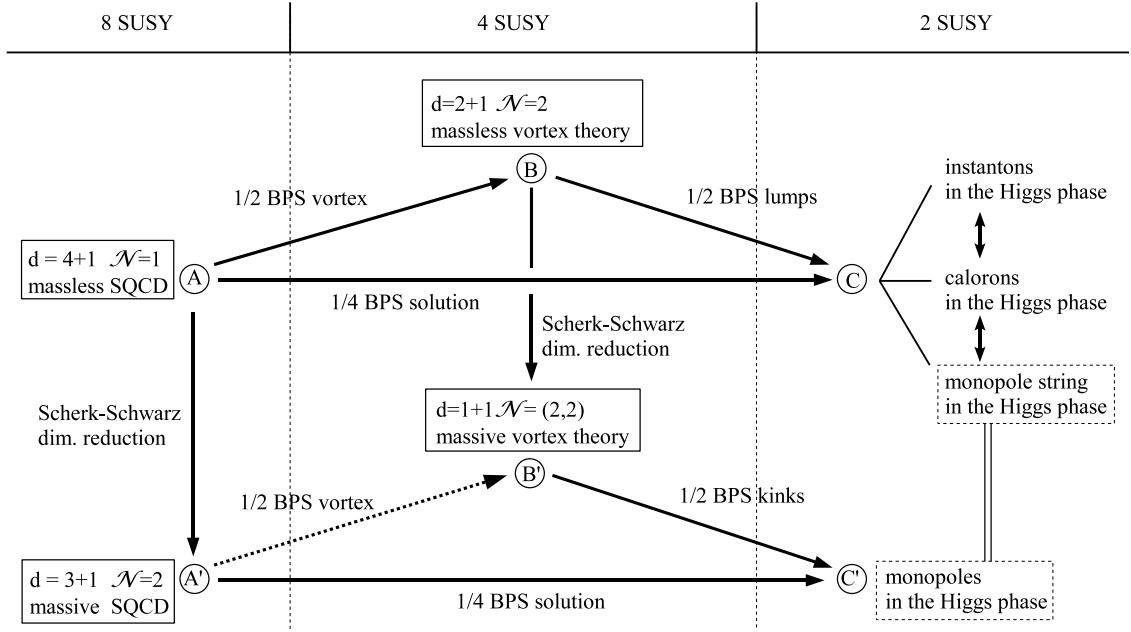}
\caption{\footnotesize 
BPS states and their relations. 
Relations discussed in this paper are denoted by 
solid lines with arrows 
and the one not discussed by a dashed line with an arrow. 
The upper triangle ($\triangle$ ABC) 
describes the $d=4+1$ gauge theory 
with massless hypermultiplets, and the lower one 
($\triangle$ A$'$B$'$C$'$)
the $d=3+1$ gauge theory with massive hypermultiplets.
The theories A and A$'$ contain eight supercharges, 
1/2 BPS vortices and their effective theories B and B$'$ 
four supercharges and 
the 1/4 BPS composite states C and C$'$ two supercharges. 
The lower theory A$'$ is obtained from 
the upper one A by the Scherk-Schwarz dimensional 
reduction (A $\to$ A$'$) preserving SUSY. 
The effective theory on a single vortex in $d=4+1$ (B) 
or $d=3+1$ (B$'$) 
is the SUSY ${\bf C}P^{N-1}$ model without or with a potential, 
respectively. 
Here the latter B$'$ coincides with the one obtained from 
the former B by the Scherk-Schwarz dimensional reduction 
(B $\to$ B$'$) preserving SUSY. 
Instantons in C and monopoles C$'$ attached by vortices 
as $1/4$ BPS states can be interpreted as 
$1/2$ BPS lumps (B$\to$C) and kinks (B$' \to$C$'$) 
in the effective theories on the vortex 
in $d=4+1$ and $d=3+1$, respectively. 
In $d=4+1$ calorons interpolates between 
instantons within a vortex 
and a monopole-string in the Higgs phase (see C). 
The monopole-string in C in $d=4+1$ 
can be dimensionally reduced to a monopole C$'$ 
in the Higgs phase in $d=3+1$. 
}
\end{center}
\end{figure}

The moduli matrix for instantons 
also contains all moduli parameters 
in all the different topological sectors.
Since the moduli matrix as a function of 
two holomorphic variables contains 
infinitely many moduli parameters, 
it is now difficult to specify all of them 
corresponding to all the solutions. 
Instead, we specify a moduli matrix for 
multiple instantons on a single vortex 
by interpreting them as $1/2$ BPS multiple lumps 
in the effective theory 
on the world-volume of the single $1/2$ BPS vortex,   
similarly to the case of monopoles 
in the Higgs phase. 
Monopoles in the Higgs phase 
can be obtained in ${\cal N}=2$ (eight SUSY) 
massive SUSY QCD (SQCD) in $d=3+1$ dimensions. 
They can be promoted to 
monopole-strings in $d=4+1$. 
The  ${\cal N}=2$ (eight SUSY) 
massive SQCD in $d=3+1$ dimensions can be 
obtained from our eight SUSY massless SQCD 
in $d=4+1$ dimensions by a Scherk-Schwarz dimensional 
reduction~\cite{Scherk:1979zr} preserving SUSY. 
It has been found that in the Coulomb phase 
there is a soliton called caloron 
that interpolates between an instanton and a 
monopole~\cite{cal1,cal2}. 
We clarify relations between 
a monopole-string and instantons in the Higgs phase 
and show that they can be obtained 
as particular limits in a wider class of 
solutions, namely calorons 
in the Higgs phase. 
Various BPS states and their relations considered 
in this paper are illustrated in 
Fig.~\ref{diagram}. 
We hope that present work opens 
a new direction in the research of instantons and monopoles.

This paper is organized as follows. 
In Sec.~\ref{bps equations} 
we derive the $1/4$ BPS equations
both by requiring the preservation of SUSY 
in the SUSY transformation laws on the fermions  
and by performing a Bogomol'nyi completion of the energy density.
We solve them and 
show that all solutions are generated by 
the moduli matrix. 
The moduli space 
with all possible boundary conditions is also clarified. 
In Sec.~\ref{instantons} we first identify the moduli matrix 
for a single vortex where 
we determine the coefficient (the K\"ahler class) 
of the K\"ahler potential  
in the effective theory on a single vortex (A$\to$B). 
Promoting moduli parameters in that matrix 
to functions of the world-volume coordinates 
we obtain the moduli matrix for a certain class of 
$1/4$ BPS states, given as 
multiple instantons inside a single vortex (A$\to$C). 
We also determine the topology of the moduli space 
for that topological sector. 
In Sec.~\ref{mono_cal} we discuss the relation 
between theories and solitons in 
$d=4+1$ (the triangle ABC) and 
$d=3+1$ (the triangle A$'$B$'$C$'$). 
It is shown that the 1/4 BPS equations and 
the projection operators  
for instantons and vortices 
in $d=4+1$ reduce to those for monopoles, vortices and walls 
in $d=3+1$. The instanton charge reduces to the monopole charge. 
We find that a particular form 
of the moduli matrix (in $d=4+1$) reproduces 
composite states made of walls, vortices and monopoles, 
uniformly distributed to one direction. 
Therefore the total moduli space of these composite states 
found in \cite{Isozumi:2004vg} has one-to-one 
correspondence with a subset of 
the total moduli space 
for composite states of vortices and instantons. 
We also clarify the relations between 
these total moduli spaces for 1/4 BPS states and 
those for $1/2$ BPS walls found in \cite{Isozumi:2004jc,INOS4} 
and for $1/2$ BPS vortices. 
Then we specify the moduli matrix 
for a monopole in the Higgs phase 
in $d=3+1$.
As a by-product we give a new way to obtain the potential in the 
effective theory on a single vortex in 
the massive theory. 
(That potential was originally discussed by 
D.~Tong~\cite{Tong:1999mg}.) 
We then discuss the calorons in the Higgs phase 
using the vortex effective theory.
Sec.~\ref{discussion} is devoted to conclusion and discussion. 
We discuss classification of all solutions of our $1/4$ BPS equations. 
There we construct another interesting solution, 
intersecting vortices 
whose intersecting point carries instanton charges. 
We thus conclude that there exist two kinds of instantons 
in the Higgs phase; the one is an instanton inside a vortex and 
the other is an instanton stuck at the intersection of vortices. 
1/4 BPS dyonic instantons are also discussed.

\section{1/4 BPS Equations and Solutions
\label{bps equations}
}
\subsection{1/4 BPS Equations for Vortices and Instantons}

We work with a $U(\NC)$ gauge theory with $N_{\rm F}$ 
massless hypermultiplets in the fundamental representation 
in $d=4+1$ dimensions\footnote{
Since we consider massless hypermultiplet, 
our consideration applies equally well for 
theories in $d=5+1$ dimensions. 
Our convention for the metric is 
$\eta_{MN}={\rm diag.}(+1, -1, \cdots, -1)$. 
} 
as the minimum dimension with 
four spacial dimensions allowing instantons (A in Fig.\ref{diagram}). 
We consider the minimum number of supersymmetry (SUSY) 
which is eight in our case. 
Since we consider massless hypermultiplet, 
we have an $SU(\NF)$ flavor symmetry. 
We consider the case of $\NF \geq \NC$.
The physical fields contained in the vector multiplet 
are a $U(\NC)$ gauge field $W_M$ $(M=0,1,\cdots,4)$, 
symplectic Majorana spinors $\lambda^i$ with $SU(2)_R$ indices $i=1,2$ 
and a real
adjoint scalar field $\Sigma$. 
The physical fields contained in the hypermultiplets are 
complex scalars $H^{irA}$ $(r=1,2,\cdots,\NC,\ A=1,2,\cdots,\NF)$
and Dirac spinors $\psi^{rA}$.
We express $\NC\times\NF$ matrix of hypermultiplets by $H^i$. 
The bosonic Lagrangian takes the form of
\be
\Lag = {\rm Tr}
\left[
- \frac{1}{2g^2}\ F_{MN}F^{MN}
+ \frac{1}{g^2}\D_M \Sigma \D^M \Sigma 
+ \D^M H^i(\D_M H^i)^\dagger 
- \frac{1}{g^2} (Y^a)^2 - H^i(H^i)^\dagger \Sigma^2
\right],
\ee
where the trace is taken over the color indices, 
$g$ is the gauge coupling constant taken common for 
$U(1)$ and $SU(\NC)$ parts of the $U(\NC)$ gauge group, 
in order to allow simple solutions later. 
The covariant derivatives and the field strength are defined by
 $\D_M \Sigma = \partial_M \Sigma + i[W_M,\Sigma]$, 
 $\D_M H^i = \partial_M H^i + iW_M H^i$ and 
 $F_{MN}=-i[\D_M,\D_N]=
 \partial_M W_N -\partial_N W_M +i[W_M, W_N]$,  
respectively. 
Here $Y^a$ are auxiliary fields of the vector multiplet 
which are determined by their equations of motion as 
\be
Y^a=\frac{g^2}{2}
\left( c^a {\bf 1}_{N_{\rm C}} 
- (\sigma^a)^j{}_{i} H^i H^{j\dagger }  \right),
\quad (a=1,2,3)
\ee
with the Pauli matrices $\sigma^a$ for $SU(2)_R$
and real parameters $c^a$ called 
the Fayet-Iliopoulos (FI) parameters.

The SUSY transformation of the fermionic fields 
are given by 
\be
\delta_\varepsilon \lambda^i
&=&\left( \frac{1}{2}\gamma^{MN} F_{MN}+\gamma^M \D_M \Sigma \right)
\varepsilon^i + iY^a(\sigma^a)^i{}_j\varepsilon^j,
\label{susy1}\\
\delta_\varepsilon \psi
&=&\sqrt{2}\left( -i\gamma^M \D_M H^i + \Sigma H^i\right)
\epsilon_{ij}\varepsilon^j.
\label{susy2}
\ee
Here the anti-symmetric tensor is defined by $\epsilon_{12} = \epsilon^{12}=1$.
The $SU(2)_R$ rotation allows us to choose the FI parameters 
as $c^a = (0,0,c>0)$ without loss of generality.
Then conditions for supersymmetric vacua are obtained as
\be
H^1(H^1)^\dagger - H^2(H^2)^\dagger = c{\bf 1}_{\NC},\quad
H^1(H^2)^\dagger = 0,\quad
\Sigma H^i = 0.
\ee
Since the non vanishing FI parameter in the first equation 
does not allow $H^i=0$ for both $i$, 
the third equation requires $\Sigma$ to vanish. 
Hence the vacua are in the Higgs branch 
with completely broken $U(\NC)$ gauge symmetry.
For $N_{\rm F} > \NC$ the moduli space of vacua is 
the cotangent bundle over 
the complex Grassmann manifold, 
$T^* G_{\NF,\NC} 
= T^* [SU(\NF)/(SU(\NF - \NC) \times SU(\NC) 
  \times U(1))]$~\cite{Grassmann}.

Recently it has been suggested 
that this model admits 
BPS states containing 
both 
non-Abelian vortices and 
instantons\cite{Hanany:2004ea}. 
The Bogomol'nyi completion for energy density 
in static configurations
can be performed as~\cite{Hanany:2004ea}
\be 
 \label{BPSbound}
{\cal E} &=& \Tr\left[
\frac{1}{2g^2}F_{mn}F_{mn} + \D_mH(\D_mH)^\dagger
+ \frac{1}{g^2}(Y^3)^2 \right]\nonumber\\
&=& 
\Tr\bigg[
\frac{1}{g^2}\left\{\left(F_{13} + F_{24} + Y^3\right)^2 
+ \left(F_{12} - F_{34}\right)^2
+ \left(F_{14} - F_{23}\right)^2
\right\} \nonumber\\
&+& 4\bar\D_z H (\bar\D_z H)^\dagger + 4\bar\D_w H (\bar\D_wH)^\dagger
- c(F_{13} + F_{24}) + \frac{1}{2g^2}F_{mn}\tilde F_{mn} 
+ \partial_mJ_m
\bigg]\nonumber\\
&\ge& \Tr\left[-c(F_{13} + F_{24}) 
+ \frac{1}{2g^2}F_{mn}\tilde F_{mn} + \partial_mJ_m\right],
\ee
where we define 
$\tilde F_{mn} \equiv (1/2)\varepsilon_{mnkl}F^{kl}$ with
$m,n \; (=1,2,3,4)$ which denote spatial indices 
for the four codimensional coordinates of solitons.
We also define two complex coordinates and the covariant derivatives 
as 
\be
 z \equiv x^1 + ix^3, \quad  
 w \equiv x^2 + ix^4, \qquad\quad
 \bar\D_z \equiv \frac{\D_1 + i\D_3}{2}, \quad
 \bar\D_w\equiv \frac{\D_2+i\D_4}{2}, 
\ee 
respectively. 
In Eq.~(\ref{BPSbound}) we have assumed 
$H^2=0$ for simplicity,~\footnote{ \label{H2=0}
If there are vortices and $H^2\not=0$, we can show 
that fields increase indefinitely away from the vortex
and energy density diverges, at least 
for simple cases of $U(1)$ gauge theory. 
}  
and we have simply denoted $H \equiv H^1$. 
Here we have also ignored 
$\Sigma$ 
because it vanishes for our $1/4$ BPS states except 
in sect.\ref{discussion} where we restore $\Sigma$ 
in order to discuss more general solution including 
dyonic instantons. 
The last line of Eq.(\ref{BPSbound}) 
gives the BPS bound for the energy density. 
Its first term counts topological charges 
for vortices in the 1-3 plane and the 2-4 plane extending to 
the 2-4 plane and the 1-3 plane, respectively,  
and the second term for the instantons.
The current $J_m$ is defined by 
$J_1
=
{\rm Re}\left(-i{\cal D}_3H H^\dagger 
\right)$,
$J_3
=
{\rm Re}\left(i{\cal D}_1H H^\dagger 
\right)$, 
and similarly for $2, 4$ directions. 
It gives a surface term which 
does not contribute to the energy of solitons 
 integrated over the entire space. 
By using the BPS equations given below, 
it can be rewritten as 
$J_m \equiv (1/2) {\cal D}_m(HH^\dagger)$. 

The BPS equations minimizing the energy density 
can be obtained from (\ref{BPSbound}) as \cite{Hanany:2004ea}:
\begin{eqnarray}
&& F_{12} = F_{34}, \quad
F_{23} = F_{14}, \quad
\bar\D_z H = 0, \quad \ \bar\D_w H = 0 
, 
\label{eq:1/4bps1}
\\
 && 
 F_{13} + F_{24} 
= - \frac{g^2}{2}\left[c{\bf 1}_{\NC} - HH^\dagger\right]
.
  \label{eq:1/4bps2}
\end{eqnarray}
The first two equations in Eq.(\ref{eq:1/4bps1}) 
give an integrability condition 
for differential operators $\bar\D_z$ and $\bar\D_w$ 
\begin{eqnarray}
[\bar\D_z,\bar\D_w] 
 =\frac{i}{4}\left[(F_{12}-F_{34})+i(F_{14}-F_{23})\right] = 0.
\end{eqnarray}
If we turn off the FI parameter $c$ and set $H=0$, 
these equations reduce 
to the self-dual equation for instantons. 
On the other hand,
if we ignore the $x^2,x^4$ $(x^1,x^3)$ dependence 
and $W_2,W_4$ $(W_1,W_3)$, 
these equations reduce to the BPS equations 
for vortices in the 1-3 (2-4) plane. 

We now show that all configurations satisfying 
the BPS equations (\ref{eq:1/4bps1}) and (\ref{eq:1/4bps2}) 
preserve $1/4$ (but not 1/8) SUSY.\footnote{
The authors in Ref.~\cite{Hanany:2004ea} suspected 
that solutions of Eqs.(\ref{eq:1/4bps1}) and 
(\ref{eq:1/4bps2}) preserve 1/8 SUSY, 
but it is not the case.
} 
To this end we introduce 
projections on the fermionic supertransformation 
parameters $\varepsilon$: it is 
specified by the subspace with positive eigenvalues 
of gamma matrices $\Gamma$ ($\Gamma^2=1$)
in the form of $\Gamma \varepsilon = \varepsilon$. 
The gamma matrices $\Gamma_{\rm v}$ for the projection 
allowing vortices in the 1-3 plane, 
$\Gamma_{\rm v'}$ for vortices in the 2-4 plane 
and $\Gamma_{\rm i}$ for instantons are given by 
\be
\Gamma_{\rm v} = - \gamma^{13}\otimes i \sigma^3,\quad
\Gamma_{\rm v'} = - \gamma^{24}\otimes i \sigma^3,\quad
\Gamma_{\rm i} = \gamma^0 \otimes {\bf 1}_2 ,\label{projection}
\ee
respectively. 
Each projection operator projects out different sets of 
four supercharges among eight supercharges, 
and therefore it is a projection for $1/2$ BPS states. 
By requiring $1/2$ SUSY specified by 
$\Gamma_{\rm v}$ $(\Gamma_{\rm v'})$
in the supertransformations (\ref{susy1}) and (\ref{susy2}) 
to be conserved, 
we obtain the BPS equations allowing 
vortices in the 1-3 (2-4) plane.
Similarly $\Gamma_{\rm i}$ 
leads to another $1/2$ BPS (selfdual) equations 
admitting instantons. 
Since a projection is defined by the subspace 
with positive eigenvalues of a gamma matrix $\Gamma$, 
two projections are compatible if and only if 
two gamma matrices commute with each other. 
In our case of vortices in the 1-3 and the 2-4 planes 
and instantons,  any two of all three gamma matrices 
$\Gamma_{\rm v}$, $\Gamma_{\rm v'}$ and $\Gamma_{\rm i}$ 
commute with each other. 
Therefore we can impose all three projections simultaneously
to preserve 1/4 SUSY. 
By requiring the supertransformation (\ref{susy1})
and (\ref{susy2}) to be conserved for the 1/4 SUSY,
we obtain the BPS equations (\ref{eq:1/4bps1}) 
and (\ref{eq:1/4bps2}) again. 
Note that any of the three projections can be derived 
from the product of the other two, 
for example $\Gamma_{\rm v'} = \Gamma_{\rm v}\Gamma_{\rm i}$.
Therefore we conclude that all solutions of
the BPS equations (\ref{eq:1/4bps1}) and (\ref{eq:1/4bps2}) 
preserve $1/4$ SUSY.

\subsection{Solutions and Their Moduli Space}
Let us solve the BPS equations (\ref{eq:1/4bps1}) 
and (\ref{eq:1/4bps2}) 
by generalizing the method introduced in \cite{Isozumi:2004jc} 
(A$\to$C in Fig.\ref{diagram}). 
The 
four equations in Eq.~(\ref{eq:1/4bps1}) 
can be formally solved as 
\be
 \bar W_z = - iS^{-1} \bar\partial_z S,\quad
 \bar W_w = - iS^{-1} \bar\partial_w S,\quad
 H = S^{-1}H_0(z,w), 
\label{eq:hyper-sol}
\ee
with $W_z$ and $W_w$ defined by
\be
 \bar W_z \equiv \frac{W_1+iW_3}{2},\ \quad \bar W_w \equiv \frac{W_2+iW_4}{2} 
\ee
and an $\NC\times\NC$ non-singular matrix function 
$S(x^m)$ is defined as a solution of the first two equations
in (\ref{eq:hyper-sol}). 
Then 
the last two equations in Eq.~(\ref{eq:1/4bps1}) 
is solved by Eq.~(\ref{eq:hyper-sol}) with 
an $\NC\times\NF$ matrix $H_0(z,w)$ 
whose components are arbitrary 
holomorphic functions with respect to $z$ and $w$. 
The matrix $H_0(z,w)$ should have rank $N_{\rm C}$ in generic 
points $(z, w)$. 
We call $H_0$ the {\it moduli matrix} 
because all moduli parameters of solutions are 
expected to be 
contained in this matrix\footnote{
In the next section, we explicitly show that 
the moduli matrix contains all the moduli parameters 
in the case of the single non-Abelian vortex.
In the presence of at least one vortex, 
the equation $H^2=0$ holds as explained in 
the footnote \ref{H2=0} 
and therefore no moduli parameters 
appear in $H^2$. 
There exists possibility such that $\Omega$ defined 
in Eq.~(\ref{omega}) below 
contains additional moduli parameters.
We have to prove the index theorem for our 1/4 BPS states 
to clarify this point.
}.
There is an important symmetry,
which we call the 
{\it world-volume symmetry}~\cite{Isozumi:2004jc}, 
defined by 
\be 
 H_0 \rightarrow H_0' = V H_0, \quad
 S \rightarrow S' = V S  \label{world-volume-tr}
\ee
with $V(z,w)$ an element of $GL(\NC,{\bf C})$ 
whose components are holomorphic with respect to 
$z$ and $w$. 
The world-volume symmetry (\ref{world-volume-tr}) 
relates sets of $(H_0,S)$ and $(H_0',S')$ which 
give the same physical quantities and defines 
an equivalence relation~\cite{Isozumi:2004jc,Isozumi:2004vg,INOS4}.
Then the total moduli space ${\cal M}_{\rm vv'i}$ 
including {\it all} topological sectors with different boundary 
conditions 
can be identified as a quotient 
of the holomorphic maps defined by
\begin{eqnarray}
&&{\cal M}_{\rm vv'i} =   
{\cal H}\backslash {\cal G}, 
\label{eq:moduli-space}
\\ 
&&{\cal G}  \equiv  
\{H_0 \ |\ {\bf C}^2 
{\longrightarrow} 
M(N_{\rm C} \times N_{\rm F}, {\bf C}), 
\bar\partial_z H_0 = \bar\partial_w H_0=0\}
\nonumber \\ 
&&{\cal H}  \equiv  \{V \ |\  {\bf C}^2 
{\longrightarrow} GL(\NC,{\bf C}),
\bar\partial_z V = \bar\partial_w V=0\},
\nonumber 
\end{eqnarray} 
where $M(N_{\rm C} \times N_{\rm F}, {\bf C})$ 
is an $N_{\rm C} \times N_{\rm F}$ complex matrix.
The dimension of this moduli space is of course 
infinite because it contains topological sectors with 
arbitrary numbers of topological charges.
By enforcing a boundary condition properly
we can obtain a topological sector with finite dimension, 
as shown in the next section. 
\footnote{
One should recall that the total moduli space (in our language)
of the sigma model instanton is 
the whole space of the holomorphic map from ${\bf C}$ 
to the target space $M$~\cite{Polyakov:1975yp,Perelomov:1987va}. 
Requiring that infinity should be mapped into a single point in $M$, 
the total moduli space is decomposed into topological sectors,  
according to the homotopy class of the map $S^2 \to M$. 
Then each topological sector contains finite number 
of moduli parameters.  \label{sigma-model-instanton}
}
All topological sectors are patched together to 
form the total moduli space.\footnote{
We have not yet clarified 
how the total moduli space is decomposed 
into different topological sectors 
in the case of instantons. 
It was, however, 
completely clarified~\cite{Isozumi:2004jc,Eto:2005wf}
in the case of the moduli space of 
the domain walls given in Eq.~(\ref{wall-moduli}), below,   
which is obtained by the dimensional reduction of this system. 
}
We would like to emphasize that 
the same thing occurs in the case of the composite states 
made of monopoles, vortices and walls~\cite{Isozumi:2004vg} 
(see Eqs.~(\ref{moduli-wvm}) and (\ref{moduli-wvm-inf}), below). 
If we put the same requirement with the footnote~\ref{sigma-model-instanton}, 
all vortices end on walls 
with perpendicular angle. 
However once ignoring such a requirement, 
we were able to obtain tilted walls where vortices 
end with angle. 
Changing boundary conditions produces new solutions. 
{\it 
Therefore considering the total moduli space 
is very important to exhaust the all solutions of BPS equations 
for composite states. }

Once $H_0$ is given, 
the matrix function $S(x^m)$ can be 
determined by the last two equations in 
Eq.~(\ref{eq:1/4bps1})
up to the $U(\NC)$ gauge transformation.
To solve it, it is useful to introduce a gauge invariant matrix 
\be
 \Omega \equiv SS^\dagger \label{omega}
\ee
which transforms as $\Omega \to V \Omega V^{\dagger}$
under the world-volume transformation (\ref{world-volume-tr}). 
Then the remaining BPS equation (\ref{eq:1/4bps2}) can be 
reexpressed in terms of $\Omega$ as 
\be
4\bar\partial_z\left(\partial_z\Omega\Omega^{-1}\right)
+ 4\bar\partial_w\left(\partial_w\Omega\Omega^{-1}\right)
= cg^2\left({\bf 1}_{\NC} - \Omega_0\Omega^{-1}\right),\quad
c\Omega_0 \equiv H_0H_0^\dagger.\label{1/4master}
\ee
We call this the master equation for our $1/4$ BPS system. 

The energy density for the $1/4$ BPS 
states consists of contributions with 
the vorticity densities $\rho_{\rm v}$ in the 1-3 plane 
and $\rho_{\rm v'}$ in the 2-4 plane, the instanton 
number density $\rho_{\rm i}$, and the 
current divergence for the correction term 
$\partial_mJ_m$ as 
\be
{\cal E}
= 
2 \pi c \left(\rho_{\rm v} + \rho_{\rm v'}
\right)
+ \frac{8\pi^2}{g^2}\rho_{\rm i}
+ \partial_m {\rm Tr} J_m
.\label{vvi_tension}
\ee
The vorticity densities $\rho_{\rm v}$ in the 1-3 
plane and $\rho_{\rm v'}$ in the 2-4 plane 
are given in terms of $\Omega$ as 
\begin{equation}
\rho_{\rm v}\equiv -\frac{1}{2\pi} {\rm Tr}F_{13} 
= \frac{1}{\pi} \bar\partial_z\partial_z\log\det\Omega, 
\quad 
\rho_{\rm v'}\equiv -\frac{1}{2\pi} {\rm Tr}F_{24} 
= \frac{1}{\pi} \bar\partial_w\partial_w\log\det\Omega, 
\label{eq:vortex_density}
\end{equation}
For finite energy configurations, $\Omega$ must 
approach $\Omega_0$ at asymptotic spacial infinity 
in the codimensions (along the direction perpendicular 
to the vortex). 
Therefore topological charge such as the vorticity 
in the 1-3 plane $\nu_{\rm v}$ 
is determined by boundary conditions 
encoded in $\Omega_0$ as 
\be
\nu_{\rm v} \equiv 
\int dx^1dx^3 \ \rho_{\rm v}
= \frac{1}{2\pi i}\oint_\infty dz\ \partial_z
\log\det\Omega_0,
\label{eq:13vorticity}
\ee
and similarly for vorticicty $\nu_{{\rm v}'}$ in the 2-4 plane.
Then the maximal power of $|z|^2$ and $|w|^2$ of the determinant of $\Omega_0$
gives the vorticity $\nu_{\rm v}$ and $\nu_{{\rm v}'}$, respectively.
On the other hand, the instanton density 
$\rho_{\rm i}$ is given in terms of $\Omega$ as 
\be
\rho_{\rm i} \equiv 
 \frac{1}{16\pi^2}{\rm Tr}(F_{mn} \tilde F_{mn}) 
=\frac{1}{\pi^2}
\Tr\left[\bar\partial_z\!\left(\partial_w\Omega\Omega^{-1}\right)
\bar\partial_w\!\left(\partial_z\Omega\Omega^{-1}\right)
- \bar\partial_z\!\left(\partial_z\Omega\Omega^{-1}\right)
\bar\partial_w\!\left(\partial_w\Omega\Omega^{-1}\right)\right]
.
\label{eq:inst-density}
\ee
To obtain the instanton number $\nu_{\rm i}$, we should just integrate 
over the density $\rho_{\rm i}$ over the Euclidean four space as
$
\nu_{\rm i}
= \int d^4x\ \rho_{\rm i}. 
\label{eq:inst-number}
$
In the case of instantons in the Higgs phase, $\Omega$ approaches 
to $\Omega_0$ at the infinity ($|z|\to\infty$) perpendicular to the 
host vortices in the 1-3 plane.
On the other hand, it does not approach to $\Omega_0$ at infinities
$(|w|\to\infty)$ along the host vortices, but
to the solution of 1/2 BPS vortices $\Omega_{\rm v}(z,\bar z)$ which
does not depend on $w,\bar w$.
Instanton charges can be calculated by evaluating
asymptotic values of $\Omega,\ \partial_w\Omega$ and $\bar\partial_w\Omega$
at infinities ($|w|\to\infty$).
This can be correctlly performed by using the Manton's effective 
action \cite{Manton:1981mp} of the host vortices as we will
show in the next section.
Finally the current divergence for the correction term 
$\partial_mJ_m$ is given in terms of $\Omega$ as 
\begin{equation}
 \partial_m {\rm Tr} J_m = 
-{2 \over g^2}\partial_m^2 
{\rm Tr}\left[\bar \partial_z(\Omega^{-1}\partial_z \Omega)+
\bar \partial_w(\Omega^{-1}\partial_w \Omega)\right]. 
\label{eq:correction_corrent}
\end{equation}

Theories with $\NF > \NC$  
are called semi-local theories, 
and vortices in these theories are called semi-local vortices \cite{semilocal}. 
In this case of $\NF > \NC$ 
we can consider the strong gauge coupling limit 
$g^2\rightarrow\infty$ in which 
the model reduces to the nonliniear sigma model
whose target space is the cotangent bundle over the 
complex Grassmann manifold, 
$T^* (G_{\NF,\NC})$~\cite{Grassmann}, 
and semi-local vortices become Grassmannian sigma-model lumps. 
In this limit the master equation (\ref{1/4master})
becomes the algebraic equation
as~\cite{Isozumi:2004jc,Isozumi:2004vg,INOS4}
\be 
 \Omega^{g^2 \to \infty} = \Omega_0 
= c^{-1} H_0 H_0^\dagger \;.
\label{eq:omega-g-infty}
\ee
Eq.~(\ref{eq:omega-g-infty}) requires the moduli matrix 
$H_0$ to have rank $N_{\rm C}$ for entire complex 
$(z, w)$ plane, in order for $\Omega$ to be invertible.
Therefore the moduli space in this limit becomes simply 
the space of all the holomorphic maps from the complex two plane 
to the complex Grassmann manifold
\begin{equation}
 {\cal M}_{\rm vv'i}^{g^2 \rightarrow\infty} 
 = \{\varphi| {\bf C}^2 \rightarrow G_{N_{\rm F},N_{\rm C}}, 
   \bar \partial_z \varphi = \bar \partial_w \varphi = 0 \}.
  \label{vvi-moduli}
\end{equation}

Let us recall that the moduli space ${\cal M}_{\rm vv'i}$ 
in Eq.~(\ref{eq:moduli-space}) 
at finite $g^2$ admits isolated points where the rank 
of $H_0$ is less than $N_{\rm C}$. 
Such isolated points correspond to 
Abrikosov-Nielsen-Olesen (ANO) vortices~\cite{ANO}
sizes of whose cores are 
of order $L_{\rm v} \sim 1/(g\sqrt{c})$.
In the infinite gauge coupling limit, the ANO vortices
tend to zero-size singular configurations of the delta function.
In the sigma models ($g^2\to\infty$) such singular configurations
give small lump singularities and are no longer points 
in the moduli space ${\cal M}_{\rm vv'i}^{g^2\to\infty}$.
In other words,
the small lump singularities in 
${\cal M}_{\rm vv'i}^{g^2 \rightarrow\infty}$ 
are blown up in the moduli space 
${\cal M}_{\rm vv'i}$ for finite gauge coupling 
by inserting the degrees of freedom of the ANO vortices. 
However not the all singularities in the moduli space 
${\cal M}_{\rm vv'i}^{g^2 \rightarrow\infty}$ 
in the strong gauge coupling limit 
are smoothed out in the moduli space ${\cal M}_{\rm vv'i}$
for finite gauge coupling. 
The moduli space ${\cal M}_{\rm vv'i}$ still has 
singularities interpreted as small instanton singularities 
as shown in the next section.

For the case of $\NF = \NC$ where we cannot take the infinite
gauge coupling limit the moduli space purely contains
degrees of freedom of the ANO vortices and instantons
as studied in the next section.

\section{Instantons in the Higgs phase
\label{instantons}
}

We have derived the master equation (\ref{1/4master}) 
of our $1/4$ BPS system and have shown 
that the total moduli space is given by 
${\cal M}_{\rm vv'i} \simeq {\cal H}\backslash {\cal G}$ 
in Eq.~(\ref{eq:moduli-space}) (A$\to$C in Fig.\ref{diagram}). 
It is, however, difficult to clarify  
what configuration each point of the moduli space gives,
because we cannot solve 
the master equation (\ref{1/4master}) in its full generality. 
In order to overcome this problem partially, 
we here restrict ourselves to consider $1/4$ BPS solutions 
which can be interpreted as $1/2$ BPS lumps (B$\to$C in Fig.\ref{diagram}) on 
the world volume of $1/2$ BPS vortices in 
the 1-3 plane (A$\to$B in Fig.\ref{diagram}).\footnote{
We cannot exhaust all $1/4$ BPS states by this method.
For example, vortices in the 2-4 plane cannot be expressed 
in the effective theory on the world-volume of the vortex. 
We will return to discuss this problem in the final Sec.\ref{discussion}.
}
Such restricted solutions constitute a moduli subspace in 
the total moduli space ${\cal M}_{\rm vv'i}$ 
defined as the space of all the holomorphic maps 
from complex plane to the vortex moduli space 
${\cal M}_{\rm v} \simeq {\bf C} \times \hat{\cal M}_{\rm v}$
(given in Eq.(\ref{modu_vortex}) below):
\be
\{ \varphi | {\bf C} \to {\cal M}_{\rm v}, \bar \partial_w \varphi = 0 \} 
\simeq {\bf C} \times \{ \varphi| {\bf C} \to \hat{\cal M}_{\rm v}, 
                       \bar \partial_w \varphi = 0\}
\subset {\cal M}_{\rm vv'i},
\label{lump_on_v}
\ee
where the factor ${\bf C}$ representing 
the center of positions of the vortices 
is factored out from the target space 
because lumps cannot wrap it.
In the case of a single vortex
the reduced moduli space 
$\hat{\cal M}_{\rm v}$ coincides with the target space of lumps
and the moduli space (\ref{lump_on_v})
reduces to that of the lump as will be shown below. The further study
is required for the case of multiple vortices.

This section consists of two subsections.
In the first subsection 
we give the effective action on 1/2 BPS vortices, 
which we call the vortex theory (B in Fig.\ref{diagram}),   
and explain a relation between the 1/4 BPS states
and the 1/2 BPS states in the vortex theory (A$\to$B$\to$C in Fig.\ref{diagram}).  
In particular we work out the vortices in 
the theory with $\NC = \NF \equiv N$ forcusing on $N = 2$, 
but not the semi-local vortices with $\NF > \NC$.
In the second subsection we find the 1/4 BPS solutions
for instantons in the Higgs phase by 
embedding the lump solution holomorphically 
into the moduli matrix for the vortex of 1/4 BPS solutions 
(A$\to$C in Fig.\ref{diagram}).

\subsection{Instantons as lumps on vortices}
Let us first work out the 1/2 BPS non-Abelian vortex (A$\to$B in Fig.\ref{diagram}).
The $1/2$ BPS equations for the vortices in the 1-3 plane
can be derived under $1/2$ SUSY condition 
with the projection defined by $\Gamma_{\rm v}$. 
They are obtained by ignoring dependence on $w$ 
in Eqs.~(\ref{eq:1/4bps1}) and (\ref{eq:1/4bps2}).
Then the master equation for vortices is also 
obtained by throwing away the $w$-dependence 
in Eq.~(\ref{1/4master}).
The moduli matrix $H_{{\rm v}0}$ 
for vortices 
does not depend on $w$ and 
is holomorphic with respect to $z$.
The total moduli space of the non-Abelian vortices is also
obtained from ${\cal M}_{\rm vv'i}$ in Eq.(\ref{eq:moduli-space}) by ignoring
$w$ dependence:
\begin{eqnarray}
&&{\cal M}_{\rm v} =  
{\cal H}_{\rm v}\backslash {\cal G}_{\rm v}, 
\label{modu_vortex}
\\ 
&&{\cal G}_{\rm v}  \equiv  
\{H_{{\rm v}0} \ |\ {\bf C} 
{\longrightarrow} 
M(N_{\rm C} \times N_{\rm F}, {\bf C}), 
\bar\partial_z H_{{\rm v}0}=0\}
\nonumber \\ 
&&{\cal H}_{\rm v}  \equiv  \{V \ |\  {\bf C}
{\longrightarrow} GL(\NC,{\bf C}),
\bar\partial_z V =0\}.
\nonumber 
\end{eqnarray} 
The effective Lagrangian using the method of Manton \cite{Manton:1981mp} is obtained by
promoting the moduli parameters $\phi^i$ in the background solutions 
with $\Omega_{\rm v}(H_0(\phi),H_0^*(\phi^*))$ 
to fields $\phi^i(x^u)$ depending on 
the world-volume coordinates $x^u$ ($u=0,2,4$) on vortices.
After a lengthy 
calculation taking the Gauss's law into account,
we find the following
effective Lagrangian on the world volume of the 
vortices~\cite{eff-theory}
in terms of $\Omega_{\rm v}$ (B in Fig.\ref{diagram}):
\be
{\cal L}_{\rm v}
&=& \int\!\! d^2x\bigg[
\delta^u \delta_u^\dagger c\log\det\Omega_{\rm v}\nonumber\\
&+& \frac{4}{g^2}\Tr\left\{
\bar\partial_z\!
\left(\delta^u\Omega_{\rm v}\Omega_{\rm v}^{-1}\right)
\delta_u^\dagger\!
\left(\partial_z\Omega_{\rm v}\Omega_{\rm v}^{-1}\right)
- \bar\partial_z\!
\left(\partial_z\Omega_{\rm v}\Omega_{\rm v}^{-1}\right)
\delta_u^\dagger\!
\left(\delta^u\Omega_{\rm v}\Omega_{\rm v}^{-1}\right)\right\}
\bigg],\label{eff-lag}
\ee
where the variation $\delta_u$ and its conjugate 
$\delta_u^\dagger$ 
act on complex moduli fields as 
$\delta_u \Omega_{\rm v}
= \sum_i\partial_u\phi^i(\delta\Omega_{\rm v}/\delta \phi^i)$ 
and 
$\delta_u^\dagger \Omega_{\rm v}
= \sum_i\partial_u \phi^{i*}
(\delta\Omega_{\rm v}/\delta \phi^{i*})$, 
respectively.

The original theory with $\NF > \NC$ 
reduces to the nonlinear sigma model 
on $T^* G_{\NF,\NC}$ 
in the strong gauge coupling limit $g^2 \to \infty$ 
as stated in the last section. 
Semilocal vortices become 1/2 BPS 
Grasmannian sigma-model lumps. 
The second term in the effective Lagrangin 
(\ref{eff-lag}) for the vortices vanishes in this limit 
and we get the K\"ahler potential 
of the effective action for the Grasmanniann lumps as 
$K_{\rm lumps} = c \int d^2x\ \log\det\Omega^{g^2\to\infty}$ 
with $\Omega^{g^2\to\infty}$ in Eq.~(\ref{eq:omega-g-infty}).  
This form of the K\"ahler potential is 
well known in the case of 
the ${\bf C}P^{\NF-1}$ lumps
corresponding to the case of $\NC=1$~\cite{Ward:1985ij,Stokoe:1986ic}.

Let us now clarify the correspondence
between
1/4 BPS states of the parent theory (A$\to$B in Fig.\ref{diagram})
and 1/2 BPS states on the vortex theory (A$\to$B$\to$C in Fig.\ref{diagram}).
To this end,
it is very important  to observe 
a relation between the effective Lagrangian (\ref{eff-lag}) of the vortex theory
and the energy density (\ref{vvi_tension}) of 1/4 BPS states
with Eqs.(\ref{eq:vortex_density}) and (\ref{eq:inst-density}).
Similarly to the last equation in Eq.~(\ref{eq:1/4bps1}), 
the $1/2$ BPS equation for lumps on the vortex theory 
is obtained as 
\be
 \bar\partial_w\phi^i = 0 . \label{BPSeq-lump}
\ee
Assuming a static solution, we obtain 
\be
\delta^u \times \delta_u^\dagger
= \partial^u \phi^i\frac{\delta}{\delta\phi^i} \times 
\partial_u \phi^{j*}\frac{\delta}{\delta\phi^{j*}}
= -2 \left(
\partial_w\phi^i\frac{\delta}{\delta\phi^i} \times 
\bar\partial_w\phi^{j*}\frac{\delta}{\delta\phi^{j*}}
+ \bar\partial_w\phi^i\frac{\delta}{\delta\phi^i} \times 
\partial_w\phi^{j*}\frac{\delta}{\delta\phi^{j*}}
\right).
\ee
The $1/2$ BPS equation (\ref{BPSeq-lump}) 
on the vortex theory implies that the variation 
$\delta^u \times \delta_u^\dagger$ 
can be identified with 
$-2\partial_w\times\bar\partial_w$ on the $1/2$ BPS states 
\be
\delta^u \times\delta_u^\dagger |\text{BPS on vortex}\rangle
= -2\partial_w\times\bar\partial_w|\text{BPS on vortex}\rangle.
\ee
Thus the effective Lagrangian (\ref{eff-lag})
evaluated on the 1/2 BPS states correctly gives the minus of the 
energy dinsity (\ref{vvi_tension}) omitting the contribution
of vortices in the 1-3 plane.
More explicitly, the first term in Eq.(\ref{eff-lag}) 
corresponds to the vortices in the 2-4 plane and the 
second term to the instantons.
This assures that instantons as $1/4$ BPS states can be identified 
as $1/2$ BPS states on the vortex theory.

To avoid inessential complications, 
let us consider the theory
in the case of $N = 2$.
The moduli space for a single vortex in this theory 
was found to be ${\bf C} \times {\bf C}P^1$ 
in Refs.~\cite{Hanany:2003hp,Auzzi:2003fs} 
where ${\bf C}$ parameterizes and  
a zero mode for broken translational symmetry 
in the two codimensions 
and ${\bf C}P^1$ for broken global $SU(2)_{\rm F}$ symmetry 
in the internal space, respectively.  
Let us first find out the moduli matrices 
for the single vortex and 
recover the previous results in Ref.\cite{Hanany:2003hp,Auzzi:2003fs}
in terms of the moduli matrices (A$\to$B in Fig.\ref{diagram}).
As was mentioned in the previous section,
$\det \Omega_0 \propto \det H_0H_0^\dagger$
for the single vortex has to be proportional 
to $|z-z_0|^2$ with $z_0$ the position of the vortex. 
We find that general moduli matrices for 
the single vortex can be transformed 
by the world-volume symmetry (\ref{world-volume-tr}) 
to either of the following two matrices:
\be
H_{{\rm v}0}^{\rm single}(z;z_0,b) \equiv \sqrt c\left(
\begin{array}{cc}
z-z_0 & 0\\
b & 1
\end{array}
\right),\quad
H_{{\rm v}0}^{'{\rm single}}(z;z_0,b') \equiv
\sqrt c
\left(
\begin{array}{cc}
1 & b'\\
0 & z-z_0
\end{array}
\right),
\label{MM_vortex1}
\ee
with $b,b'\in {\bf C}$.
These two matrices can be transformed to each other 
with the relation $b'=1/b$ 
by a world-volume transformation (\ref{world-volume-tr})
\footnote{
$
H_{{\rm v}0}^{'{\rm single}}(z;z_0,b'=1/b) = 
\left(
\begin{array}{cc}
0 & 1/b\\
-b & z-z_0
\end{array}
\right)H_{{\rm v}0}^{\rm single}(z;z_0,b)$.
}
except for specific points $b=0$ and $b'=0$.
Clearly, the moduli space of the single vortex can be identified as
${\bf C} \times S^2 \simeq {\bf C} \times {\bf C}P^1$.
More explicitly, ${\bf C}$ and $S^2 \simeq {\bf C}P^1$
are covered by $z_0$ and 
by two patches $b$ and $b'$ in Eq.~(\ref{MM_vortex1}), respectively.
The moduli parameter $b$ can be identified as 
the orientational moduli parameter
which is associated with 
the spontaneously broken $SU(2)_{\rm F}$ flavor symmetry\cite{Hanany:2003hp,Auzzi:2003fs}.
In fact, the above moduli matrix can be rederived 
from that with $b=0$ by
acting $U\in SU(2)_{\rm F}$ combined 
with a world-volume transformation $V_U$:
\be
H_{{\rm v}0}^{\rm single}(z;z_0,b)
 = V_U H_{{\rm v}0}^{\rm single}(z;z_0,0) U,
\ee
with
\be
U \equiv
 \left(
 \begin{array}{cc}
 \phi_1 & \phi_2\\
 -\phi_2^* & \phi_1^*
 \end{array}
 \right),\quad
V_U \equiv
 \left(
 \begin{array}{cc}
 \phi_1^* & -\phi_2 (z-z_0)\\
 0 & 1/\phi_1^*
 \end{array}
 \right).
\label{eq:su2moduli}
\ee
Here $|\phi_1|^2 + |\phi_2|^2 = 1$ and we have 
identified $b = -\phi_2^*/\phi_1^*$.
Note that $H_{{\rm v}0}^{\rm single}(z;z_0,0)$ breaks $SU(2)_{\rm F}$ into $U(1)_{\rm F}$,
so the orientational moduli space is found to be 
${\bf C}P^1 \simeq SU(2)_{\rm F}/U(1)_{\rm F}$ parametrized by 
homogenious coordinates $\phi_1$ and $\phi_2$ 
or an inhomogenious coordinate $b$ or $b'$.
$H^{\rm single}_{{\rm v}0}(z;z_0,b)$ 
is the most general for a single vortex 
in the sense that it contains all solutions with 
a single vortex.
This is consistent with the results in Ref.\cite{Hanany:2003hp}, where
the real dimension of the moduli space of the non-Abelian $k$ vortices
was obtained as $2kN$ by making use of the index theorem.  

Let us next solve the master equation for the single vortex with the
moduli matrix (\ref{MM_vortex1}).
To do this, we first recall the case of $N=1$, 
where we obtain the 
well-known ANO vortex. 
In our formulation, the ANO vortex is given by $\Omega_\star$ 
satisfying 
\be
4\bar\partial_z\partial_z\log\Omega_\star 
= cg^2\left(1 - |z-z_0|^2\Omega_\star^{-1}\right).
\ee
Returning to the $N=2$ case, let us first 
take the diagonal moduli matrix $H_{{\rm v}0}^{\rm single}(z;z_0,0)$.  
The solution of the master equation 
(\ref{1/4master}) 
for the $1/2$ BPS vortex with this moduli matrix 
$H_{{\rm v}0}^{\rm single}(z;z_0,0)$ is obtained by embedding the ANO vortex 
solution $\Omega_\star$ 
as $\Omega_{\rm v}\big|_{b=0}= {\rm diag.} (\Omega_\star,1)$ 
\cite{Hanany:2003hp}.
The 
solutions corresponding to the general 
moduli matrix $H_{{\rm v}0}$ in Eq.~(\ref{MM_vortex1}) 
can be obtained by 
using the world-volume transformation $V_U$ in 
Eq.(\ref{eq:su2moduli}) as
\be
\Omega_{\rm v} 
= V_U (\Omega_{\rm v}\big|_{b=0}) V_U^\dagger 
= \left(
\begin{array}{cc}
\frac{\Omega_\star + |b|^2|z-z_0|^2}{1 + |b|^2} & \bar b(z-z_0)\\
b(\bar z - \bar z_0) & 1 + |b|^2
\end{array}
\right). \label{sol}
\ee

We now reach the place where 
the effective theory of the single vortex can be exactly obtained
(B in Fig.\ref{diagram}).
This can be achieved by 
promoting the moduli parameters $z_0$ and $b$ 
in the solution (\ref{sol}) into fields on the vortex 
world-volume $z_0(t,w,\bar w)$ and $b(t,w,\bar w)$ and 
by plugging them into the effective Lagrangian (\ref{eff-lag}) \cite{Manton:1981mp}.
We thus find
the K\"ahler potential 
with the coefficient of the K\"ahler metric (K\"ahler class) 
$4\pi/g^2$
for the full moduli fields 
$z_0(t,w,\bar w)$ and $b(t,w,\bar w)$ :
\be
K_{\rm v} 
= c\pi|z_0|^2 + \frac{4\pi}{g^2}\log\left(1 + |b|^2\right).
\label{kahler}
\ee
The first term comes from the first term of the effective Lagrangian
(\ref{eff-lag}) and the second term from the second term of (\ref{eff-lag})
corresponding to the instantons.
This Lagrangian with the K\"ahler class 
$4\pi / g^2$ was also determined 
in \cite{Hanany:2003hp} by the brane configuration 
and in \cite{Shifman:2004dr} 
by using the $1/4$ BPS states with 
monopoles in the Higgs phase.

Following the prescription given in the introduction,
next we consider the $1/2$ BPS lumps in 
the effective theory on the 2+1 dimensional world volume 
of the vortex (B$\to$C in Fig.\ref{diagram}).
The BPS equation (\ref{BPSeq-lump}) can be solved 
for $k$-lumps using
rational functions of degree $k$~\cite{Polyakov:1975yp,Ward:1985ij} 
as
\be
b(w) = \frac{P_k(w)}{\alpha P_k(w) + aQ_{k-1}(w)},\label{k-lumps}
\ee
with
\be
P_k(w) \equiv \prod_{i=1}^{k}(w-p_i),\quad
Q_{k-1}(w) 
\equiv
\prod_{j=1}^{k-1} (w- q_j).
\ee
The moduli parameters $\{p_1,p_2,\cdots,p_k\}$ 
have one to one correspondence with the positions 
of the $k$-lumps in the host vortex,
$a$ with the total size of the configurations
and $\{q_1,q_2,\cdots,q_{k-1}\}$ with the relative sizes of the $k$-lumps.
The remaining modulus $\alpha$ 
parametrizes ${\bf C}P^1$ at the 
bounday (the infinity) of $w$ 
since $b(w) \to 1/\alpha$ as $|w| \to \infty$.
Especially,  $\{p_i,a,q_j\}$ can precisely be identified with
positions and sizes of $k$-lumps when $\alpha=0$.\footnote{
For $\alpha \neq 0$ we should redefine these parameters 
to describe the physical positions and sizes. 
For the case of $k=1$ 
see Eq.~(\ref{eq:pos-size}).
}
Notice that zeros of the denominater in Eq.~(\ref{k-lumps})
are  not true singularities 
but mere coordinate singularities.
This is an artifact caused by the fact that
$b$ is an inhomogenious coordinate of 
the ${\bf C}P^1$ manifold.
Namely the corresponding configurations are smooth and 
continuous at these coordinate singularities. 
On the other hand, the point $a=0$ and 
the points $p_i =q_j$ 
are true singularities of the moduli space of 
the lumps
since the degree of the solution (\ref{k-lumps}) decreases and
the corresponding configurations become singular.
These singularities are called small lump singularities.

As was expected,
the mass of $k$-lumps precisely agrees with 
that of the $k$-instantons, namely $8\pi^2k/g^2$.
This mass comes from the second term in Eq.(\ref{kahler})
which originally corresponds to the instanton charges as was shown in the 
previous section.
We thus can identify  
the $1/4$ BPS instantons in the original theory in $d=4+1$ 
dimensions (A$\to$C in Fig.\ref{diagram})
as the $1/2$ BPS lumps in the effective theory on the 
$d=2+1$ dimensional world-volume of the vortex (A$\to$B$\to$C in Fig.\ref{diagram}).

\bigskip
Returning to the vortex, 
orientational moduli space for 
spontaneous symmetry breaking by 
the single non-Abelian vortex 
for the case of $N>2$ 
was shown to be
$SU(N)/[SU(N-1)\times U(1)] \simeq 
{\bf C}P^{N-1}$~\cite{Hanany:2003hp,Auzzi:2003fs,Eto:2004ii}.
One of the patch for moduli space of the single vortex is given by
\be
H_{{\rm v}0} = \sqrt c
\left(
\begin{array}{ccccc}
z-z_0   & 0 & 0 & \cdots &0\\
b_1     & 1 & 0 & \cdots & 0\\
b_2     & 0 & 1 & \ddots & \vdots\\
\vdots  & \vdots & \ddots & \ddots & 0\\
b_{N-1} & 0 & \cdots & 0 & 1
\end{array}
\right) ,\label{moduli_U(N)}
\ee
with $b_i$ complex parameters.
There exist other $N-1$ patchs for ${\bf C}P^{N-1}$ 
given through the world-volume transfomation (\ref{world-volume-tr}).
There exist $N$ complex moduli parameters $z_0$ and $b_i$. 
Here $z_0$ is the position of the vortex
and $b_i$ are orientational moduli parametrizing ${\bf C}P^{N-1}$. 
The K\"ahler potential for the orientational moduli parameters $b_i$
can be determined up to the constant
factor as $K \propto \log(1+\sum |b_i|^2)$ 
by discussing only symmetry.
The factor can be precisely determined without exact solutions 
or any calculations 
by recognizing an equivalence between 
the mass of the $1/4$ BPS objects
in the original theory and 
the $1/2$ BPS objects in the vortex theory. 
Then we get
\be
 K_{\rm v} = \pi c|z_0|^2 + \frac{4\pi}{g^2}\log\left(1 + \sum_{i=1}^{N-1}|b_i|^2\right).
\ee
The multi-lump solution for the ${\bf C}P^{N-1}$ model 
is also known \cite{Stokoe:1986ic}.

\subsection{1/4 BPS solutions of the instantons in the Higgs phase}

The aim of this subsection is to specify the moduli matrix $H_0(z,w)$ 
for the instantons in the Higgs phase as 
the 1/4 BPS states (A$\to$C in Fig.\ref{diagram}), 
which have been found to be the 1/2 BPS lumps 
on the vortex theory in the previous subsection (A$\to$B$\to$C in Fig.\ref{diagram}).
We will also specify 
the moduli space of the instantons 
in the Higgs phase.
Our basic strategy is to replace 
the moduli parameter $b$ in the moduli matrix 
$H_{{\rm v}0}^{\rm single}(z;z_0,b)$ in Eq.~(\ref{MM_vortex1}) 
for a single vortex 
by the lump solution $b(w)$ in Eq.~(\ref{k-lumps}):\footnote{
The exact relation between these matrices is given 
in Eq.~(\ref{relation}), below.
}
\be
H_0(z,w) \sim H_{{\rm v}0}^{\rm single}(z;z_0,b(w))
= \sqrt c
\left(
\begin{array}{cc}
z - z_0 & 0\\
\frac{P_k}{\alpha P_k + aQ_{k-1}} & 1
\end{array}
\right).
\label{1/2_to_1/4}
\ee
Although this procedure is very simple, 
there exists a technical complication; 
a solution $b(w)$ is not holomorphic 
at some points in $w$ where $b(w)$ diverges, 
whereas all components in the moduli matrix $H_0(z,w)$ 
have to be holomorphic with respect to both $z$ and $w$ 
at any point $(z,w)\in{\bf C}^2$ 
to cover the whole solutions consistently. 
This can be overcome by noting that 
the lump solution $b(w)$ is given in 
an inhomogeneous coordinate $b$ on ${\mathbf C}P^1$. 
Therefore we should 
transform the moduli matrix 
$H_{{\rm v}0}^{\rm single}(z;z_0,b(w))$
written in the inhomogeneous cooordinate $b$ 
into the one in homogenious coordinates. 
This can be achieved by
\be
H_0(z,w)
= \sqrt c\left(
\begin{array}{cc}
(z-z_0)A_{k-1}(w) & (z-z_0)(\alpha A_{k-1}(w) + aB_{k-2}(w))\\
P_k(w) & \alpha P_k(w) + aQ_{k-1}(w)
\end{array}
\right) , \label{k-instantons}
\ee
with $A_{k-1}$ and $B_{k-2}$ being 
the polynomial functions of order $k-1$ and $k-2$ in $w$, given by 
\be
A_{k-1}(w) = \sum_{i=1}^{k} \frac{1}{Q_{k-1}(p_i)}
\prod_{i' (\neq i)=1}^{k}\left(\frac{w-p_{i'}}{p_i-p_{i'}}\right),\ \ 
B_{k-2}(w) = \sum_{j=1}^{k-1} \frac{-1}{P_k(q_j)}
\prod_{j' (\neq j)=1}^{k-1}\left(
\frac{w-q_{j'}}{q_j-q_{j'}}\right).
\label{AB}
\ee
These $A_{k-1}(w)$ and $B_{k-2}(w)$ have been 
uniquely determined by the condition\footnote{
We can consider the polynomial functions 
$A_{k'}$ of order $k'$ and  $B_{k'-1}$ of order $k'-1$ for $k'>k-1$.
However, we can always set $k'=k-1$ 
by use of the world-volume transformation without loss of generality.
}
\be
 A_{k-1}Q_{k-1} - B_{k-2}P_k = 1 
\label{aq-bp}
\ee
requiring that
the vorticity of the solution should coincide with
the one in Eq.~(\ref{1/2_to_1/4}), namely that 
the solutions should have a single vortex in the 1-3 plane
and no vortices in the 2-4 plane. 
Now the relation between the right hand side of Eq.~(\ref{1/2_to_1/4}) 
and Eq.~(\ref{k-instantons})
is shown to be
\be
H_0(z,w)
= V(P_k(w),Q_{k-1}(w))
\ H_{{\rm v}0}^{\rm single}(z;z_0,b(w)), \label{relation}
\ee
with the matrix $V(P_k,Q_{k-1})$ defined by
\be
V(P_k,Q_{k-1}) \equiv \left(
\begin{array}{cc}
\dfrac{a}{\alpha P_k + aQ_{k-1}} & (z-z_0)(\alpha A_{k-1} + aB_{k-2})\\
0 & \alpha P_k + aQ_{k-1}
\end{array}
\right).
\ee
This matrix is a valid world-volume 
transformation (\ref{world-volume-tr}) only in a 
particular region of $w$ 
with $\alpha P_k + aQ_{k-1}$ non-zero,
since it has a singularity in $w$. 
Although $V(P_k,Q_{k-1})$ is not
a valid world-volume transformation (\ref{world-volume-tr}) 
because of singularities in $w$,
it is needed to
obtain the regular moduli matrix 
(\ref{k-instantons})  by compensating singularities in 
$H_{{\rm v}0}^{\rm single}(z;z_0,b(w))$.

Next let us examine the moduli parameters of the $k$-instantons
in the Higgs phase in detail.
No new parameters appear in $A_{k-1}$ and $B_{k-2}$, and therefore 
the configuration of $k$-instantons in the Higgs phase 
has the $2k+2$ complex moduli parameters
$(z_0,\{p_i\},\{q_j\},a,\alpha)$. 
Here $z_0$ is the position of the single vortex on the 1-3 plane.
As was mentioned in the first of this section, this decouples with
other moduli paramters. 
So the moduli space of this configuration
can be simply written as
\be
{\cal M}^{k\text{-instantons}} \simeq
{\bf C} \times {\cal M}^{k\text{-lumps}}
\simeq {\bf C} \times 
   \{\varphi|{\bf C} \to \hat{\cal M}^{1\text{-vortex}}, 
             \bar \partial_w \varphi = 0\}.
\ee
Note that this decoupling property of 
$z_0$ from $(\{p_i\},\{q_j\},a,\alpha)$
can also be read from the K\"ahler potential (\ref{kahler}). 
From the discussion given in the previous subsection,
we realize that $\{p_i\}$ correspond to 
the positions of $k$-instantons inside the vortex, 
$a$ to the total size and the orientation of the configurations
and $\{q_j\}$ to the relative sizes and 
orientations of the instantons.
It is very interesting to observe 
that the {\it small lump singuralities} 
with $a=0$ or $p_i = q_j$ 
in Eq.~(\ref{k-lumps}) are now interpreted as 
the {\it small instanton singuralities} in the Higgs phase.
In fact, in the limit with $a$ tending to zero
the rank of the moduli matrix
(\ref{k-instantons}) reduces by one and its determinant vanishes.
Then the point $a \to 0$ is singular in the moduli space.
On the other hand, the small lump singuralities coming
from $p_i = q_j$ in Eq.~(\ref{k-lumps}) appear as the
divergences of $1/P_k$ and $1/Q_{k-1}$ in $A_{k-1}$ and $B_{k-2}$ 
in Eq.(\ref{AB}).\footnote{In Eq.~(\ref{AB}) 
these also 
appear to diverge when $p_i = p_{i'}$ for $i\neq i'$ and
$q_j = q_{j'}$ for $j\neq j'$, respectively, but it is not the case;
We can show that the factors $p_i - p_{i'}$ and 
$q_j - q_{j'}$ in denominators 
are always cancelled with numerators after the summation.
Hence the points 
$p_i = p_{i'}$ and $q_i = q_{j'}$ 
are not singular in the moduli space.
} 
The remaining parameter $\alpha$ parametrizes ${\bf C}P^1$ 
similarly to the lump solutions. 
In summary we find 
$z_0 \in {\bf C}$, $p_i \in {\bf C}$, 
$a \in {\bf C}^* \equiv {\bf C} - \{0\} \simeq 
{\bf R}\times S^1$,  
$q_j \in 
{\bf C} - \{p_1,p_2,\cdots,p_k\}$ 
and $\alpha \in {\bf C}P^1$.

\begin{figure}[t]
\begin{center}
\includegraphics[width=8cm]{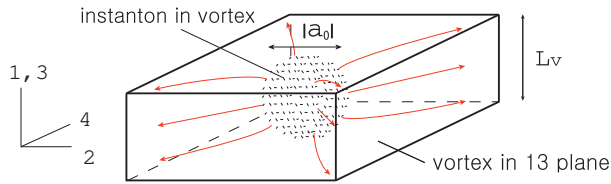}
\caption{Single instanton in the Higgs phase. The size of the vortex
is given by $L_{\rm v} \sim 1/g\sqrt c$.}
\label{iihp}
\end{center}
\end{figure}

\medskip
Now we discuss the simplest case of 
a single instanton $(k=1)$ with $A_0 = 1$ and $B_{-1}=0$ 
in more detail.
Then we have
\be
 b(w) = \frac{w-p_1}{\alpha(w-p_1) + a}
  \quad \Rightarrow \quad
 H_0^{\text{1-instanton}} = \sqrt c
 \left(
  \begin{array}{cc}
   z-z_0 & \alpha(z-z_0)\\
   w-p_1 & \alpha(w-p_1) + a
  \end{array}
 \right).
\label{eq:1vortex-moduli-matrix}
\ee
To clarify the physical significance of these 
four complex moduli parameters $z_0,p_1,a,\alpha$, 
let us transform the moduli matrix in 
Eq.(\ref{eq:1vortex-moduli-matrix})
into that with $\alpha = 0$
by the $SU(2)_{\rm F}$ rotation. 
This can be perfomed by
choosing $\phi_2=-\alpha\phi_1^*$ of $U$ in 
Eq.(\ref{eq:su2moduli}) and factor out the 
world-volume symmetry in Eq.(\ref{world-volume-tr}).
Then we get the physical
position $p_0$ and the size 
$|a_0|$ of the instanton in the vortex 
\be
 p_0 = p_1-\frac{\alpha^*}{1+|\alpha|^2}a, 
  \quad
 |a_0| = 
 \frac{|a|}{1+|\alpha|^2} .
\label{eq:pos-size}
\ee
These are invariant under the $SU(2)_{\rm F}$ transformation.
We illustrate this configuration in four Euclidean space 
schematically in Fig.~\ref{iihp}.

Let us discuss the global structure (topology)
of the the moduli space ${\cal M}^{1\text{-instanton}}$ of
one instanton.
The moduli matrix in one
patch $H_0'$ 
corresponding to the second one in Eq.~(\ref{MM_vortex1}) 
is related to that in
the other patch  $H_0$ 
in Eq.(\ref{eq:1vortex-moduli-matrix})
by a world volume transformation $V$
as
\be
 H_0^{'\text{1-instanton}}
 \equiv \sqrt c
 \left(
  \begin{array}{cc}
   \alpha'(w-p'_1) + a' & w-p'_1\\
   \alpha'(z-z_0) & z-z_0
  \end{array}
 \right)
=V H_0^{\text{1-instanton}} ,
\quad
V = 
 \left(
  \begin{array}{cc}
           0 & 1/\alpha \\
   1/\alpha & 0
  \end{array}
 \right),
\label{eq:rel-vortex-moduli-matrix}
\ee
with the following relation between coordinates
in two patches
\be
 \alpha' = \frac{1}{\alpha},
  \quad
 a' = -\frac{a}{\alpha^2},
\quad
 p'_1 = p_1 -\frac{a}{\alpha}.
\ee
Here both $\alpha$ and $\alpha'$ are the patches
of the standard inhomogeneous coordinates
of the ${\mathbf C}P^1$ and they are enough to
cover the whole manifold.
We see that $p_1$ and $a$ transform
in the union of the two patches $\alpha$ and $\alpha'$.
We find that
$a$ requires a nontrivial transition function $-1/\alpha^2$
between two patches, showing that
it is a tangent vector as
a fiber on the ${\mathbf C}P^1$.
However, instead of $p_1$ we can use the coordinate $p_0$
in Eq.~(\ref{eq:pos-size}) which is
an invariant global coordinate for two patches,
indicating that the space ${\bf C}$
parametrized by $p_0$ is
a direct product to the ${\mathbf C}P^1$.
Therefore we obtain the topology of the moduli
space of one $U(2)$ instanton in the Higgs phase as
\be
 (z_0, p_0, a, \alpha) \in
{\bf C} \times
{\bf C} \times ({\bf C}^* \ltimes {\bf C}P^1)
  \simeq {\cal M}^{1\text{-instanton}}, \label{single-inst}
\ee
with $F \ltimes B$ denoting a fiber bundle over a base space
$B$ with a fiber $F$.
More precisely ${\bf C}^* \ltimes {\bf C}P^1$
is the tangent bundle without zero section.\footnote{
\label{non-commutative}
It is interesting to observe that 
the moduli space (\ref{single-inst}) with zero section added 
is homeomorphic to
that of single $U(2)$ non-commutative  instanton, 
${\bf C}^2 \times T^* {\bf C}P^1$ \cite{NS}. 
}

Here we make a comment on (non-)normalizability of zero modes.
Some zero modes corresponding to these moduli parameters 
in (\ref{single-inst})
are not normalizable under four dimensional integration 
over the all codimensions. 
For instance the modulus $z_0$ for the position of the vortex is 
normalizable in two dimensions perpendicular to the vortex
but is apparently non-normalizable in four dimensions. 
The modulus $\alpha$ parametrizes the boundary condition of 
the sigma model instanton in the effective theory on the vortex.
It is also non-normalizable in the effective theory 
and therefore in the original theory. 
Nevertheless we emphasize that 
all of the moduli parameters in 
(\ref{single-inst}) are needed 
to determine the configuration of this composite state, 
and that dynamics of the composite state 
is described by these parameters.\footnote{
This phenomenon of non-normalizable modes 
is a commonplace issue in composite solitons, and 
has been observed in the case of the wall junction \cite{Ito:2000zf}. 
Let us explain the inevitability of non-normalizable 
modes in composite solitons by taking the junction 
as the simplest example. 
The $1/4$ BPS junction can be formed if three or more 
half-infinite $1/2$ BPS walls meet at a 
junction. 
Nambu-Goldstone (NG) fermions necessarily arise 
corresponding to the broken $3/4$ of supercharges. 
One might hope that these NG fermions are localized 
around the junction and are normalizable in the 
co-dimension two plane of the junction of walls. 
One can show that they cannot be normalizable as follows. 
Take any one of the constituent walls. 
Those NG fermions corresponding to the supercharges 
broken by that wall have support on the wall, which 
extends to infinity along the wall. 
Hence they are non-normalizable as modes on the 
co-dimension two composite soliton (junction of walls). 
We can choose the remaining NG modes which do not have 
support on that particular wall. 
However, these NG fermions correspond to 
supercharges broken by at least one of the other walls. 
Then they have to have support along the walls 
which break the supercharge. 
Consequently all of the NG fermions should have 
support infinitely extending along at least one of 
the walls, and are non-normalizable. 
One can easily recognize that this feature of 
non-normalizable modes is a usual phenomenon of 
composite solitons, and care should be excercised 
when we discuss effective theories on the 
composite solitons. 
}

\medskip

For the case of $N>2$ 
we can specify the moduli matrix 
$H_0 (z,w)$ for a particular class of $1/4$ BPS solutions 
identified as $1/2$ BPS states in the vortex theory, 
similarly to the case of $N=2$. 
We could also obtain the moduli matrices and the moduli space 
for $1/4$ BPS states 
corresponding to the $U(N)$ instantons
in the Higgs phase by repeating the same discussion.

\section{Monopoles and calorons in the Higgs phase
\label{mono_cal}
}
Recently, $1/4$ BPS states of the monopoles in the Higgs phase
were studied 
\cite{Tong:2003pz}--\cite{Kneipp} 
in $3+1$ dimensional massive 
SQCD 
with non-degenerate 
masses 
for hypermultiplets. 
Unlike the case of the massless model
(or the massive model in which all the mass parameters are degenerate),
the $SU(\NF)$ flavor symmetry 
is explicitly broken 
to $U(1)^{\NF-1}$ in the massive case.
Therefore vortices in the model with non-degenerate masses
are essentially Abelian (ANO) vortices,
and moduli fields corresponding to the orientational moduli paramters 
\cite{Auzzi:2003em,Hanany:2004ea,Eto:2004ii}
are not exactly massless moduli in the effective theory of vortices.
It has been found that the effective theory of the  non-Abelian vortices
can be constructed 
in the case where the mass difference is very smaller than the FI paramter
$|\Delta m| \ll \hat g\sqrt{\hat c}$~\cite{Tong:1999mg,Tong:2003pz}.
Here
$\hat g$ and $\hat c$ are the gauge coupling and
the FI parameter in 3+1 dimensions, respectively.
The effective theory has a potential $\hat V\sim k^2$ 
where $k$ is a Killing vector on the moduli space of the non-Abelian vortices
in the massless model\cite{Tong:1999mg,Tong:2003pz}.
In the following of this section
we call the vortex theory with the potential {\it massive} vortex theory
and the vortex theory without any potential {\it massless} vortex theory.

It has been shown in Ref.\cite{Tong:2003pz} that
the 1/4 BPS state of the monopoles in the Higgs phase can be realized as the
1/2 BPS kinks in the massive vortex theory (A$'\!\to$B$'\!\to$C$'$ in Fig.\ref{diagram}).
In this section we will find the 1/4 BPS solution corresponding 
to the monopoles in the Higgs phase directly. 
Namely, we specify the moduli matrix
for the 1/4 BPS state of the monopoles in the Higgs phase 
(A$'\!\to$C$'$ in Fig.\ref{diagram}).
To achieve this, we find that it is very useful to promote the 3+1 dimensional
massive theory to the 4+1 dimensional 
massless theory (A$'$\!$\to$A in Fig.\ref{diagram}).
By this procedure a monopole in 3+1 dimensions becomes a monopole-string
in 4+1 dimensions.
This 4+1 dimensional point of view (the triangle ABC in Fig.\ref{diagram})
not only gives a nice realization of the 
monopoles but also leads to calorons in the Higgs phase which
interporate between the instantons and the monopoles in the Higgs phase.

\subsection{Walls, vortices and monopoles revisited
\label{mono}
}

The four-dimensional massive model with non-degenerate masses 
\be
M \equiv {\rm diag.}(m_1,m_2,\cdots,m_{\NF}),
\ee
for hypermultiplets can be 
derived from our five-dimensional massless SQCD
\footnote{
In Ref.\cite{Isozumi:2004vg}, we studied the {\it massive} 
SQCD in $d=4+1$ dimensions. 
This can be derived from the six-dimensional {\it massless} 
SQCD by the Scherk-Schwarz dimensional reduction, in exactly 
the same manner. 
}
by performing the Scherk-Schwarz (SS) 
dimensional reduction~\cite{Scherk:1979zr} (A$\to$A$'$ in Fig.\ref{diagram}),
in which the fifth direction $x^4$ 
is compactified on $S^1$ with radius $R$ 
using a twisted boundary condition
\be
H(x^\mu,x^4 + 2\pi R) = H(x^\mu,x^4)e^{i2\pi RM}, 
\label{twist}
\ee
with $0 \le m_A < 1/R$. 
If we ignore the infinite towers of the Kaluza-Klein modes, 
we have  the lightest mass field $\hat H(x^\mu)$ 
as a function of the four-dimensional spacetime coordinates 
\be
H(x^\mu,x^4) = \frac{1}{\sqrt{2\pi R}} \hat H(x^\mu) e^{iMx^4},
\label{ss}
\ee
with $\mu =0,1,2,3$. 
Other fields neutral under the $SU(\NF)$ flavor symmetry are of the form
\be
W_\mu(x^\mu,x^4) = W_\mu(x^\mu),\quad
\Sigma(x^\mu,x^4) - iW_4(x^\mu,x^4) = 
\Sigma(x^\mu) + i\hat\Sigma(x^\mu).
\ee
The $1/4$ BPS equations (we have ignored $H^2$ and $\Sigma$)
in (\ref{eq:1/4bps1}) and (\ref{eq:1/4bps2}) 
reduce to those in four 
dimensions~\cite{Isozumi:2004vg}
\be
&&F_{12} = -\D_3\hat\Sigma,\ 
F_{23} = -\D_1\hat\Sigma,\ 
\bar\D_z \hat H = 0,\ 
\D_2 \hat H = \hat HM - \hat\Sigma \hat H,
\label{1/4bps_4dim1}\\
&&F_{13} - \D_2\hat\Sigma 
= -\frac{\hat g^2}{2}\left(\hat c{\bf 1}_{\NC} - \hat H\hat H^\dagger\right).
\label{1/4bps_4dim2}
\ee
Here the gauge coupling $\hat g$ and
the FI paramter $\hat c$ in four dimensions are given by
\be
\frac{1}{\hat g^2} \equiv \int^{2\pi R}_0 dx^4\ \frac{1}{g^2} = \frac{2\pi R}{g^2},
\qquad
\hat c \equiv \int^{2\pi R}_0 dx^4\ c = 2\pi R c,
\ee
respectively. 
It was known that these BPS equations admit walls, 
vortices and monopoles as $1/4$ BPS states 
\cite{Tong:2003pz,Isozumi:2004vg} (A$'\!\to$C$'$ in Fig.\ref{diagram}).
The supercharges preserved by the above BPS 
equations are summarized
in the Table \ref{supercharge}:
\begin{table}[htbp]
\begin{center}
\caption{{\small 
Gamma matrices of the
projection operators for $1/4$ BPS states
in 4 and 5 dimensions.
Here we define $\gamma^5 \equiv -i\gamma^4 = i\gamma^0\gamma^1\gamma^2\gamma^3$.
In our previous paper \cite{Isozumi:2004vg},
we used another set of gamma matrices related by the 
redefinition $\gamma^m \to -\gamma^mi\gamma^5$, 
and chose the wall profile along $x^3$ instead of the 
present choice of $x^2$.
}}
	\label{supercharge}\ \\
\begin{tabular}{ccc}
	\begin{tabular}{c|c|c}
	$d=4+1$ & world-volume & $\Gamma$\\
	\hline
	vortex & 0,2,4 & $-\gamma^{13}\otimes i\sigma^3$\\
	vortex & 0,1,3 & $-\gamma^{24}\otimes i\sigma^3$\\
	instanton & 0 & $\gamma^{0}\otimes {\bf 1}_2$
	\end{tabular}
	&&
	\begin{tabular}{c|c|c}
	$d=3+1$ & world-volume & $\Gamma$\\
	\hline
	vortex & 0,2 & $-\gamma^{13}\otimes i\sigma^3$\\
	wall & 0,1,3 & $-i\gamma^{2}\gamma^5\otimes i\sigma^3$\\
	monopole & 0 & $\gamma^{0}\otimes {\bf 1}_2$
	\end{tabular}
\end{tabular}
	\end{center}
\end{table}\ \\
The Table \ref{supercharge} shows that 
vortices in the 1-3 plane, vortices in the 2-4 plane 
and instantons in five dimensions are the BPS states 
with the conserved supercharge specified by the 
same projection as vortices in the 1-3 plane, 
walls transverse to the $x^2$-direction 
and monopoles in four dimensions, respectively.
Therefore after the SS dimensional 
reduction, these BPS solitons in five dimensions 
reduce to the respective BPS solitons 
in four dimensions. 
In fact, the instanton charge coincides with 
the monopole charge 
under the SS dimensional reduction as 
\be
\frac{1}{2g^2}\int d^3x \int^{2\pi R}_0dx^4\ 
\Tr(F_{mn}\tilde F^{mn})
= \frac{2}{\hat g^2} \int d^3x\ \vec\partial\cdot
\Tr(^*\!\vec F\hat\Sigma).
\ee

The BPS equations in Eqs.(\ref{1/4bps_4dim1}) and 
(\ref{1/4bps_4dim2}) in 3+1 dimensional massive theory
have been solved in terms of the moduli matrix
of the system~\cite{Isozumi:2004vg}.
Especially all the exact solutions were obtained in the strong
gauge coupling limit $\hat g^2 \to \infty$ 
in the semilocal case with $\NF >\NC$.
Here we reconsider Eqs.(\ref{1/4bps_4dim1}) and 
(\ref{1/4bps_4dim2}) in general case of $\NF\ge\NC$ 
from the five-dimensional point of view.
For that purpose, let us consider a restricted sector
of the moduli space
which is specified by the moduli matrix $H_0(z,w)$
in the form of 
\be
H_0(z,w) = \frac{1}{\sqrt{2\pi R}}\hat H_0(z) e^{Mw},
\label{restrict_to_wvm}
\ee
where an $\NC\times\NF$ matrix $\hat H_0(z)$ 
does not depend on $w,\bar w$, and is
holomorphic with respect to $z$.
The matrix $\hat H_0(z)$ should have rank $\NC$ in generic points of $z$
(namely, apart from isolated points).
Note that this restriction (\ref{restrict_to_wvm})
is up to the world-volume transformation (\ref{world-volume-tr}).
For the restricted moduli matrix given above the ``source'' $\Omega_0$ of the 
master equation (\ref{1/4master}) is independent of the $x^4$-coordinate
$
\Omega_0(x^M) 
= \hat c^{-1} \hat H_0 e^{2Mx^2} \hat H_0^\dagger
\equiv \hat\Omega_0(x^\mu).
$
Then
the solution $\Omega$ of Eq.(\ref{1/4master}) is also independendent of
the $x^4$-coordinate:
$
S(x^M) = \hat S(x^\mu)
$
and
$
\Omega(x^M) = SS^\dagger = \hat S \hat S^\dagger \equiv \hat \Omega(x^\mu).
$
At this stage, the master equation (\ref{1/4master}) reduces to
\be
4\bar\partial_z\left(\partial_z\hat\Omega\hat\Omega^{-1}\right)
+ \partial_2\left(\partial_2\hat\Omega\hat\Omega^{-1}\right)
= \hat c \hat g^2\left({\bf 1}_{\NC} - \hat\Omega_0\hat\Omega^{-1}\right).
\label{master_wvm}
\ee
Their solution (\ref{eq:hyper-sol}) can also be rewritten as follows
\be
H 
=  \frac{1}{\sqrt{2\pi R}}
\hat S^{-1}\hat H_0e^{Mw},\quad
\bar W_z = -i\hat S^{-1}\bar\partial_z \hat S,\quad
W_2 - i\hat \Sigma = -i\hat S^{-1}\partial_2 \hat S.
\label{sol_wvm1}
\ee
Notice that the above solution automatically satisfies the condition
of the SS dimensional reduction (\ref{ss}) if we identify
\be
\hat H(x^\mu) = \hat S^{-1}(x^\mu)\hat H_0(z) e^{Mx^2}.
\label{sol_wvm2}
\ee
The master equation (\ref{master_wvm}) and its solutions (\ref{sol_wvm1})
and (\ref{sol_wvm2})
completely agree with
those for the 1/4 BPS states containing walls, 
vortices and monopoles \cite{Isozumi:2004vg}.
Therefore the restriction (\ref{restrict_to_wvm}) 
to the form of the moduli matrix
gives a map from 1/4 BPS
solutions in 3+1 dimensions to those in 4+1 dimensions
(C$\to$C$'$ in Fig.\ref{diagram}).

We now realize that all the 1/4 BPS states of the walls,
vortices and monopoles in the massive SQCD \cite{Isozumi:2004vg}
have one to one correspondence with those in the 
restricted sector (\ref{restrict_to_wvm})
of our 1/4 BPS states (vortices and instantons) 
in the five-dimensional massless SQCD.
The moduli space  of the former
can also be understood from the five-dimensional point of view as
\be
&&{\cal M}_{\rm wvm} = 
{\cal H}_{\rm wvm}\backslash {\cal G}_{\rm wvm},\\
&&{\cal G}_{\rm wvm}  \equiv  
\{\hat H_0 \ |\ {\bf C}
{\longrightarrow} 
M(N_{\rm C} \times N_{\rm F}, {\bf C}),   
\bar\partial_z \hat H_0 =0\},\nonumber\\
&&{\cal H}_{\rm wvm}  \equiv  \{V \ |\  {\bf C}
{\longrightarrow} GL(\NC,{\bf C}),
\bar\partial_z  V = 0\} ,\nonumber  \label{moduli-wvm}
\ee
where $\hat H_0(z)$ must have the maximal rank $\NC$ 
in generic $z$ except for several points.
It is interesting to observe that this total moduli space 
agree with that for the non-Abelian vortices in Eq.(\ref{modu_vortex}),
$\cal{M}_{\rm wvm} \simeq {\cal M}_{\rm v}$, although
the former is for 1/4 BPS states and the latter is for 1/2 BPS states.
In the strong gauge coupling limit
the moduli space 
${\cal M}_{\rm wvm}$
becomes 
\be
{\cal M}_{\rm wvm}^{g^2 \rightarrow\infty} 
 = \{\varphi| {\bf C} \rightarrow G_{N_{\rm F},N_{\rm C}}, 
   \bar \partial_z \varphi = 0 \},  \label{moduli-wvm-inf}
\ee
in the case of $\NF > \NC$. 
This coincides with the moduli space of 
the Grassmannian sigma-model lumps \cite{Perelomov:1987va} 
which can be classified by $\pi_2(G_{N_{\rm F},N_{\rm C}}) = {\bf Z}$ 
if we compactify ${\bf C}$ to ${\bf C}P^1$.

Moreover, if we push forward this descent relation from ${\cal M}_{\rm vv'i}$ 
to ${\cal M}_{\rm wvm}$, we arrive at solutions of the non-Abelian walls and
their moduli space ${\cal M}_{\rm w}$ which
have been extensively studied in Ref.\cite{Isozumi:2004jc,EINOOST}.
To achieve this, we ignore $z$ dependence in ${\cal M}_{\rm wvm}$:
\be
&&{\cal M}_{\rm w} = 
{\cal H}_{\rm w}\backslash {\cal G}_{\rm w}
\simeq G_{\NF,\NC},\\
&&{\cal G}_{\rm w}  \equiv  
\{ \tilde H_{0} \ |\ {\bf C}^0 
{\longrightarrow} 
 M(N_{\rm C} \times N_{\rm F}, {\bf C}) , {\rm rank} \NC  \}
= \{ M(N_{\rm C} \times N_{\rm F}, {\bf C}) ,{\rm rank} \NC \},\nonumber\\
&&{\cal H}_{\rm w}  \equiv  \{V \ |\  {\bf C}^0
{\longrightarrow} GL(\NC,{\bf C})\} = GL(\NC,{\bf C}),\nonumber
   \label{wall-moduli}
\ee
where ${\bf C}^0$ is just a point. 
The condition on the constant matrix $\tilde H_{0}$ 
to have the maximal rank $\NC$ has been deduced from 
the condition on $H_0 (z,w)$ or $\hat H_{0}(z)$ 
in generic points of $z$ or $(z,w)$ 
in the case of instantons or monopoles, respectively.
It comes from the fact that the moduli matrix 
must have rank $N_{\rm C}$ in the vacuum. 
It is interesting that 
in the strong gauge coupling $(\NF>\NC)$ the total moduli 
space is unchanged
\be
 {\cal M}_{\rm w}^{g^2\rightarrow\infty} \simeq {\cal M}_{\rm w},
\ee
unlike the case of other solitons because 
the moduli matrix $\tilde H_{0}$ is a constant matrix here.

We thus have found that solutions of our BPS equations (\ref{eq:1/4bps1})
and (\ref{eq:1/4bps2}) can give all four kinds of solitons:
walls, vortices, monopoles and instantons. 
The relations between their total moduli spaces are given by 
\be
{\cal M}_{\rm w}
\subset
{\cal M}_{\rm wvm}
\left(
\simeq
{\cal M}_{\rm v}
\right)
\subset
{\cal M}_{\rm vv'i}.
\ee

\subsection{Monopoles in the Higgs phase}

Let us next 
find the solution of Eqs.(\ref{1/4bps_4dim1}) and (\ref{1/4bps_4dim2}) 
(A$'\!\to$C$'$ in Fig.\ref{diagram})
corresponding to one monopole 
in the Higgs phase, namely a single monopole attatched to a vortex.
To be precise, we restrict ourselves into the simplest case with 
$\NC = \NF \equiv N = 2$ in the following of this section.
As was explained in Sec.~\ref{instantons}, 
the most general moduli matrix
containing a single vortex can be written in the form of 
\be
\hat H_0(z) = \sqrt{\hat c} \left(
\begin{array}{cc}
z-z_0 & 0\\
b_c & 1
\end{array}
\right),
\label{mono_in_higgs}
\ee
where $b_c$ is a constant complex paramter.
Notice that this is the same form with the moduli matrix (\ref{MM_vortex1}) 
generating only a single vortex 
in the massless theory.
However in the massive theory the moduli matrix 
(\ref{mono_in_higgs}) gives not only a vortex but also
a monopole,
where $b_c$ gives the position and the phase of a monopole
inside the vortex,
as will be shown below. The difference 
between the massless theory and the massive theory appears as 
the factor $e^{M x^2}$ in Eq.~(\ref{sol_wvm2})
which is absent in Eq.(\ref{eq:hyper-sol}).
In the massless limit $M \to 0$, 
the moduli matrix (\ref{mono_in_higgs}) gives a single vortex only
as expected.

Similarly to instantons in the Higgs phase,
we can calculate charge of the monopole in the Higgs phase
in terms of the massive vortex theory.
To this end, let us recall the following two facts.
One is that the instantons can be realized as lumps in the 
massless vortex theory with the K\"ahler potential (\ref{kahler}) 
(A$\to$B$\to$C in Fig.\ref{diagram}).
The other is that 1/4 BPS states in the four-dimensional massive theory
can be obtained by the restriction (\ref{restrict_to_wvm}) 
on 1/4 BPS states in the
five-dimensional massless theory (C$\to$C$'$ in Fig.\ref{diagram}).
Combining these facts together, we naturally arrive at a notion of
the SS dimensional reduction for the vortex theory (B$\to$B$'$ in Fig.\ref{diagram}).
In order to achieve this,
we should first find what the action of the SS dimensional reduction to
the vortex theory is.
In terms of the moduli matrix $H_0(z,w)$ the above
twisted boundary condition (\ref{twist})
can be translated in terms of the moduli matrix
as $H_0(z,w+2\pi iR) = VH_0(z,w)e^{2\pi iRM}$
where $V(z,w)$ is an element of world-volume transformation (\ref{world-volume-tr}).
This naturally induces the following twisted boundary condition
on the moduli fields $z_0(t,x^2,x^4)$ and $b(t,x^2,x^4)$ in the moduli matrix
$H_{{\rm v}0}^{\rm single}(z;z_0,b)$ in the effective theory of the 
host vortex 
\be
H_{{\rm v}0}^{\rm single}
\left(
	z;z_0(t,x^2,x^4+2\pi R),b(t,x^2,x^4+2\pi R)
\right)\nonumber\\
= V
H_{{\rm v}0}^{\rm single}
\left(
	z;z_0(t,x^2,x^4),b(t,x^2,x^4)
\right)
e^{i2\pi RM},
\ee
with $V = e^{-2\pi i RM}$.
Hence we can identify the twisted boundary condition to the 
moduli fields as
$z_0(t,x^2,x^4+2\pi R) = z_0(t,x^2,x^4)$ and 
$
b(t,x^2,x^4+2\pi R) = e^{i2\pi R\Delta m} b(t,x^2,x^4)
$
with $\Delta m \equiv m_1 - m_2$.
We thus get the action of the SS dimensional reduction for $z_0$ and $b$ as
\be
z_0(t,x^2,x^4) = \hat z_0(t,x^2),\quad
b(t,x^2,x^4) = e^{i\Delta m x^4} \hat b(t,x^2).
\label{SS_b}
\ee
Plugging  (\ref{SS_b}) into the effective Lagrangian with the
K\"ahler potential (\ref{kahler})
of
the 2+1 dimensional massless vortex 
theory,
we can obtain the effective 
Lagrangian for the massive vortex thoery after integrating over $x^4$
(B$\to$B$'$ in Fig.\ref{diagram}). 
The resulting 1+1 dimensional massive vortex theory
consists of 
the scalar potential $\hat V_{\rm v}$ 
arising from the kinetic term in extra dimension $x^4$ 
and 
the K\"ahler potential $\hat K_{\rm v}$, given by
\be
\hat K_{\rm v} 
= \hat c \pi |\hat z_0|^2 + \frac{4\pi}{\hat g^2} \log\left(1+|\hat b|^2\right),\quad
\hat V_{\rm v} 
= \frac{4\pi}{\hat g^2}
\frac{\Delta m^2|\hat b|^2}{\left(1 + |\hat b|^2\right)^2}. 
\label{massive}
\ee
These exactly agree with the results in Ref.\cite{Tong:2003pz}
including the K\"ahler class of the K\"ahler potential and
the coefficient of the scalar potential.

The scalar potential in Eq.(\ref{massive}) admits two discrete 
SUSY vacua at $\hat b = 0,\infty$.
Thus $1/2$ BPS states on the vortex theory
become kinks (B$'\!\to$C$'$ in Fig.\ref{diagram}). In fact,
the $1/2$ BPS equation (\ref{BPSeq-lump}) reduces to
\be
\partial_2\hat b - \Delta m \hat b = 0,
\label{BPSeq-kink}
\ee
and the solution interpolating between $\hat b=0$ and $\hat b=\infty$
is found to be
\be
\hat b(t, x^2) = e^{-i\theta} e^{\Delta m (x^2-x^2_0)},
\label{eq:kink}
\ee
where $e^{-i\theta}$ is a phase and
a real parameter $x^2_0$ is a position of the kink \cite{kinks}.
Although this configuration exponentially grows, the energy density 
is localized around $x^2=x^2_0$
\be
{\cal E} = \frac{4\pi}{g^2} \frac{\Delta m^2}{\cosh^2\Delta m (x^2-x^2_0)},
\label{ene_kink}
\ee
implying usual wall profile in suitable coordinates.
The mass of the kink can be obtained by integrating this over the $x^2$-coordinate
and we get
$4\pi\Delta m/\hat g^2$. 
This coincides with the mass of a monopole in the Coulomb phase\footnote{
The symmetry breaking $SU(2)_{\rm F} \to U(1)_{\rm F}$ is given by
$\Delta m$ in the Higgs phase,
and by the vacuum expectation value (VEV) of the adjoint scalar in
the case of the Coulomb phase ('t Hooft-Polyakov monopole).
In fact, by replacing $\Delta m$ by
the VEV of the adjoint scalar, we correctly reproduce 
the mass of the monopole in the Coulomb phase.}.
Thus monopoles in the Higgs phase can be seen as the
$1/2$ BPS kinks in the vortex theory\cite{Tong:2003pz}.

Before closing this subsection, let us clarify the relation between
$b_c$ in the moduli matrix (\ref{mono_in_higgs}) and
$\theta$ and $x^2_0$ 
in the 1/2 BPS kink solution (\ref{eq:kink}).
For that purpose the five-dimensional point of view
(triangle ABC in Fig.\ref{diagram})
gives us a
very nice picture.
Taking Eq.(\ref{SS_b}) into account, the monopole(kink) solution (\ref{eq:kink})
is understood in the massless vortex theory (\ref{kahler}) as
((B$'\!\to$C$'$)$\to$(B$\to$C) in Fig.\ref{diagram})
\be
b(t,x^2,x^4) = e^{i\Delta m x^4} \hat b(t,x^2) 
= e^{- (\Delta m x^2_0+i\theta)}e^{\Delta m w }.
\label{mono_string}
\ee
Note that the K\"ahler potential (\ref{kahler}) with 
this solution substituted 
is independent of $x^4$ because it
is a function of
$|b|=e^{\Delta m (x^2-x^2_0)}$.
Therefore the energy density of this configuration 
extends along
the $x^4$-axis to infinity.
Then this configuration is understood as a 1/4 BPS state of the 
monopole-string in the Higgs phase.
Let us next find the moduli matrix $H_0(z,w)$ in five dimensions corresponding to
the moduli matrix given in Eq.(\ref{mono_in_higgs}) for
the monopole in the Higgs phase in four dimensions.
This can be obtained from the first equation in Eq.(\ref{sol_wvm1}) with
$S(x^M) = \hat S(x^\mu)$ as ((A$'\!\to$C$'$)$\to$(A$\to$C) in Fig.\ref{diagram})
\be
H_0(z,w) = \sqrt c \left(
\begin{array}{cc}
z - z_0 & 0\\
b(w) & 1
\end{array}
\right)
 = \sqrt c\ 
V\left(
\begin{array}{cc}
z-z_0 & 0\\
b_c & 1
\end{array}
\right)
e^{Mw},\quad
V = e^{-Mw}.
\ee
We thus find $b(w) = b_c e^{\Delta mw}$.
Comparing this with the kink solution in (\ref{mono_string}),
the complex parameter $b_c$ can be identified 
as ((A$\to$C) = (A$\to$B$\to$C) in Fig.\ref{diagram})
\be
b_c = e^{-(\Delta mx_0^2+i\theta)}.
\ee

Hence, we conclude that the 1/4 BPS moduli matrix (\ref{mono_in_higgs})
describes a monopole in the Higgs phase and the complex parameter 
$b_c$ therein is the position and the phase of the monopole.
In the massless limit $\Delta m \to 0$ with $\Delta m x^2_0$ fixed,
$b_c$ becomes the orientational moduli of the non-Abelian
vortex
since
the $SU(2)_{\rm F}$ flavor symmetry, which 
is explicitly broken to $U(1)_{\rm F}$
by non-zero $\Delta m$,
is restored when $\Delta m = 0$.

\subsection{Calorons in the Higgs phase
\label{cal}
}

In the Coulomb phase (unbroken gauge symmetry)
there is well known way to get ordinary monopole solution
independent of $x^4$ from instanton solutions\cite{cal1,cal2}.
Range the instantons with equal size along the $x^4$-axis periodically.
After takeing the limit where the
size parameter of instantons goes to infinity the configuration
becomes one BPS monopole-string solution extending 
to the $x^4$-axis\cite{cal1,cal2}.
In the Higgs phase, instantons, a monopole-string and calorons which
interpolate between instantons and the monopole-string can be understood
in terms of the deformations of the lump solutions in the vortex theory.
In this subsection we will concentrate on 1/2 BPS states in
the 2+1 dimensional massless
vortex theory with the K\"ahler potential
(\ref{kahler}) (B$\to$C in Fig.\ref{diagram}).

Let us first examine the monopole-string solution (\ref{mono_string})
as the sigma model lump in the massless vortex theory in more detail.
We first note that the solution $b=b_ce^{\Delta m w}$ 
in Eq.(\ref{mono_string}) is a 1/2 BPS state in the vortex theory
since this is holomorphic in $w$ and is a solution of the BPS equation 
(\ref{BPSeq-lump}) in the vortex theory.
Although this solution has one co-dimension in the vortex theory, 
this is not a domain wall which
is a topological soliton supported by the homotopy group $\pi_0$
because there is no scalar potential here.
Rather, we should realize this soution as
a topological object which consists of an infinite number of
1/2 BPS lumps supported by the homotopy group $\pi_2$.
To see this, we decompose the solution as 
$b = b_c e^{\Delta m x^2} e^{i\Delta m x^4}$.
Then it is clear that 
a strip $\sigma_{\rm l}^{(k)} = \{(x^2,x^4)\ |\ 
x^2 \in (-\infty,\infty),\ x^4 \in [2\pi k/\Delta m,2\pi (k+1)/\Delta m]\}$
is mapped to the ${\bf C}P^1$ manifold once by this configuration.
Then  the solution $b=b_ce^{\Delta mw}$ 
has infinite winding number $\pi_2({\bf C}P^1) = \infty$.
Hence  this can be realized as 
topological object which has an infinite number of lump charge.
The energy density of the solution is the same form as that in Eq.(\ref{ene_kink}).
If we integrate this in a strip $\sigma^{(k)}_{\rm m} = \{(x^2,x^4)\ |\ 
x^2 \in (-\infty,\infty),\ x^4 \in [2\pi kR,2\pi (k+1)R]\}$
($R$ is a compactification radius 
associated with the SS dimensional reduction), we find that the 
tension of the solution in the strip $\sigma_{\rm m}^{(k)}$ coincides with 
to the mass of the monopole $4\pi\Delta m/\hat g^2$. 
So this solution is suitable to be called the monopole-string.

Let us next consider 1/4 BPS calorons in the Higgs phase.
\be
b(w) = \left( 1 + \theta \right) e^{\mu(w-w_0)} - \theta,\qquad
\theta\equiv \frac{1}{a\mu}.
\label{wall_to_lump}
\ee
Here, $\mu$ and $a$ are arbitrary real parameters with mass dimension one
and minus one respectively.
\begin{figure}[t]
\begin{center}
\begin{tabular}{cccc}
\includegraphics[width=3.5cm]{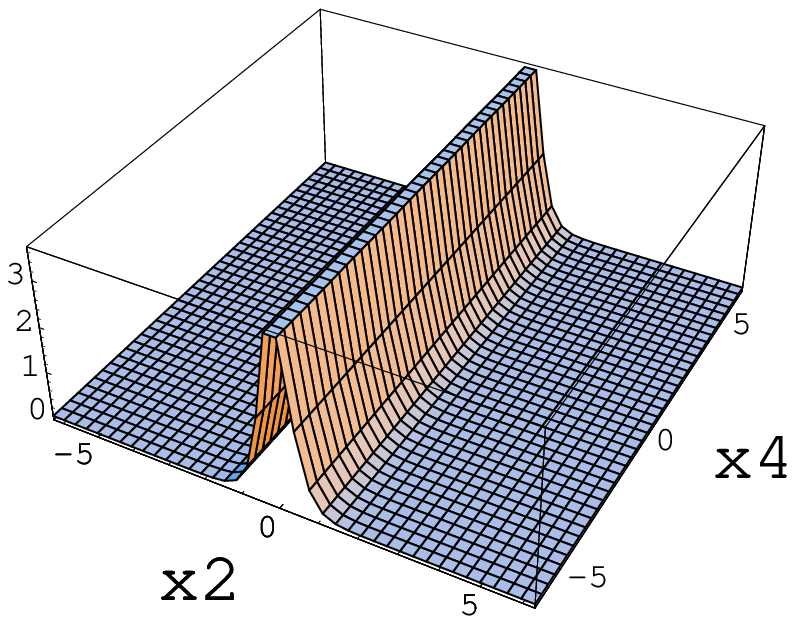}
&\includegraphics[width=3.5cm]{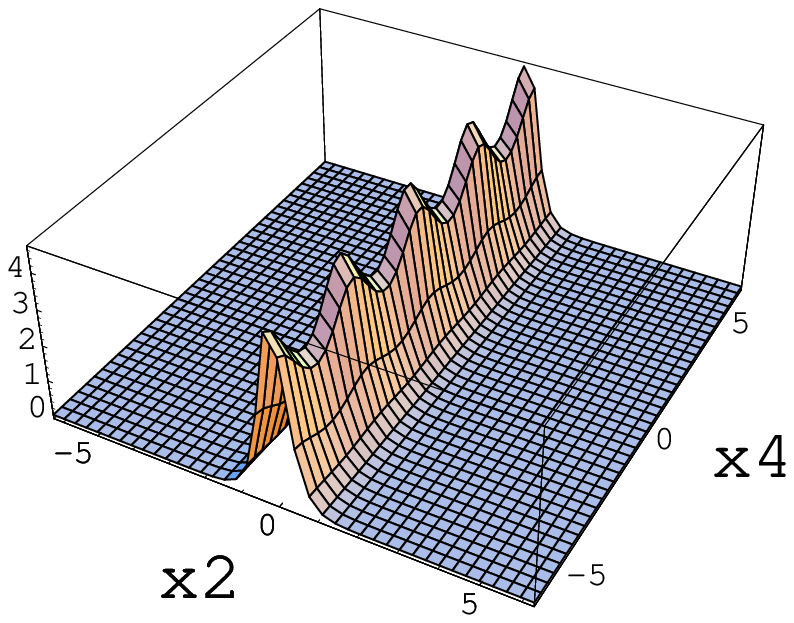}
&\includegraphics[width=3.5cm]{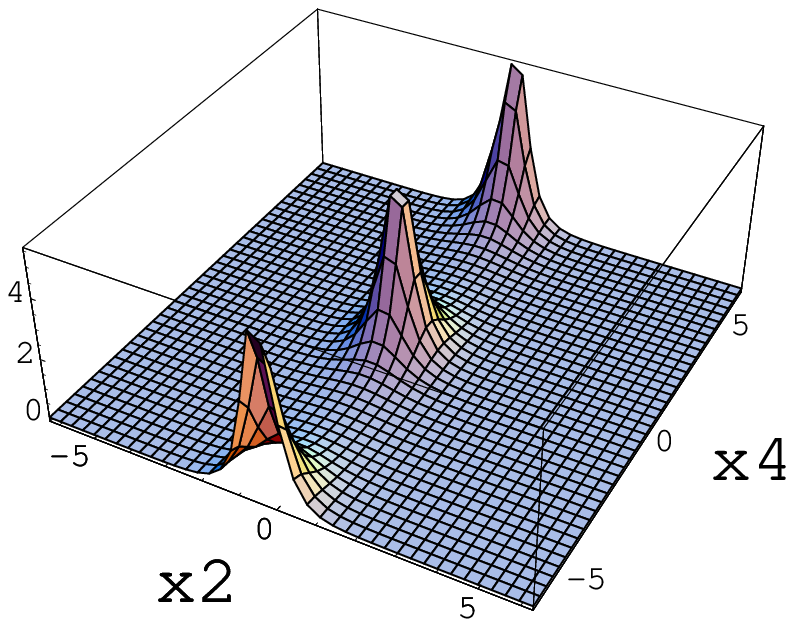}
&\includegraphics[width=3.5cm]{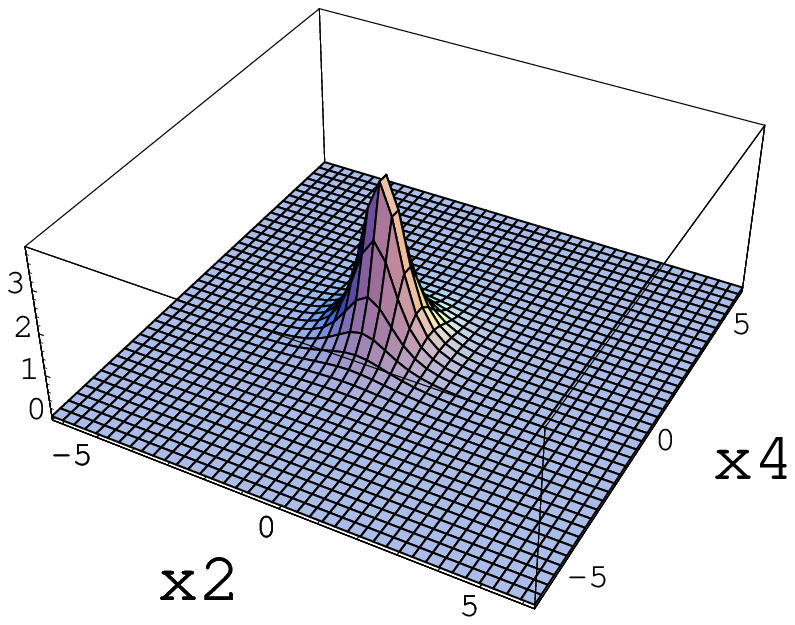}\\
(a) monopole-string & (b) caloron & (c) caloron 
& (d) instanton\\
$\mu=2,\ \theta=5\times10^{-4}$ & $\mu=2,\ \theta=0.1$ 
& $\mu=1,\ \theta=1$ & $\mu=0.2,\ \theta=5$
\end{tabular}
\end{center}
\caption{Energy density of the calorons in terms of 
the vortex theory.}
\label{caloron}
\end{figure}
Similarly to the ordinary calorons 
in the Coulomb phase\cite{cal1,cal2},
the calorons in the Higgs phase can be continuously deformed the monopole-string
or instantons.
In fact, in the limit $\theta\rightarrow0$ with $\mu$ fixed, 
this solution reduces to 
the 1/4 BPS monopole-string $b\rightarrow e^{\mu(w-w_0)}$
as shown in Fig.\ref{caloron}(a). 
There is an another limit $\theta \to \infty$ 
with $a$ fixed. 
In this limit this solution reduces
to 
$1/4$ BPS states of an instanton
in the Higgs phase 
as shown in Fig.\ref{caloron}(d). 
For general $\theta$, we find periodic lump solutions inside a vortex
which
can be understood as the 1/4 BPS caloron
as shown in Fig.\ref{caloron}(b) and (c). 
The parameter $a$ is the size of the instanton
and $1/\mu$ is the period of the caloron. 

In view of this solution, we can guess that the monopole 
strings with large instanton charges in the compactified theory 
are unstable as follows. 
If we compactify the $x^4$-direction with radius $R$,
the mass of the monopole-string solution $b=e^{\mu(w-w_0)}$ is 
$8\pi^2R\mu/g^2$. 
To be precise, let us represent $\mu = (k+\tau)/R$
with $k\in {\bf Z}$ and $\tau \in [0,1)$. 
Then the above mass of the monopole-string can be rewritten as
\be
\frac{8\pi^2}{g^2}k + \frac{4\pi}{\hat g^2}\Delta m,\quad
\left(\frac{\tau}{R} \equiv \Delta m\right).
\ee
This mass corresponds to the mass of 
$k$-instantons and a monopole with 
a ``fractional" instanton charge $\tau$. 
Then the monopole-string with mass $8\pi^2R\mu/g^2$ can be
decomposed into these solitons by continuous 
deformation like (\ref{wall_to_lump}). 
Since the aggregate of the decomposed solitons has larger 
entropy, the monopole-string with instanton charge greater 
than unity is unstable and may decay into instantons and 
a monopole-string with the fractional instanton charge, 
if this system is put at finite tempertures.

\section{Conclusion and Discussion
\label{discussion}
}

We have solved $1/4$ BPS equations for composite states 
made of instantons and vortices. 
We have shown that all solutions are generated by 
the moduli matrix which is a holomorphic 
function of $z=x^1 + i x^3$ and $w = x^2 + i x^4$. 
The moduli matrix contains all solutions 
with different boundary conditions and/or 
different topological charges. 
As a first step toward 
the complete classification of all solutions,
we have specified the moduli matrix for 
$1/4$ BPS states which can be interpreted as 
lumps on a single vortex. 
Small instanton singularities have been 
shown to correspond to small lump singularities. 
We have determined the moduli space for a single instanton 
in the Higgs phase to be the direct product of 
${\bf C}^2$ and the tangent bundle over ${\bf C}P^1$
without zero section. 
We have clarified the relations between the moduli spaces
of 1/4 BPS states for vortices and instantons, of 1/4 BPS states
for walls, vortices and monopoles,
of 1/2 BPS vortices, and of 1/2 BPS walls.
We also have constructed calorons
in the Higgs phase 
which interpolate 
between instantons and a monopole-string 
in the Higgs phase.

We did not exhaust all solutions in this paper: 
our moduli matrix contains 
more varieties of solutions. 
The complete classification of all solutions 
is a very important open problem. 
Let us discuss this issue. 
First of all we could consider multiple vortices 
as host solitons. 
However the moduli matrix for multiple vortices 
is not available yet. 
It is now in progress to specify the moduli parameters 
in the moduli matrix, 
and therefore we have to wait for the completion of that work \cite{eff-theory}
to discuss the multiple lumps on multiple vortices. 

Second, if we do not restrict ourselves to 
solitons which can be understood 
as lumps on vortices, 
we can 
obtain more varieties of solitons. 
Intersection of two or more vortices 
cannot be understood as solitons 
in the effective theory on a host vortex, 
because the energy of such solitons diverges 
in the effective theory in general. 
Instead, we can directly construct solutions 
of intersecting vortices as follows.\footnote{
Similar $1/4$ BPS states of intersecting vortices 
were discussed in \cite{Naganuma:2001pu}.
}  
In the same model with $N=2$, 
the following two moduli matrices 
give configurations with 
$\nu_{\rm v}=k_z (\ge 0)$ vortices in the 1-3 plane and 
$\nu_{\rm v'}=k_w (\ge 0)$ vortices 
in the 2-4 plane 
\be
H_0 =
\left(
\begin{array}{cc}
z^{k_z} & 0\\
0 & w^{k_w}
\end{array}
\right), \quad \quad 
H_0 =
\left(
\begin{array}{cc}
z^{k_z} w^{k_w} & 0 \\
              0 & 1
\end{array}
\right).
\ee
The vortices intersect at a point
for both cases.
It is, however, a trivial intersection for the former case, 
and they carry no instanton charge.
On the other hand, 
they intersect non-trivially for the latter case, 
and the intersecting point carries the instanton charge 
$\nu_{\rm i} = - k_z k_w$. 
In this case the instantons give a negative energy contribution. 
However, there is no inconsistency, 
since the total energy including vortices is always 
positive.\footnote{
A similar situation occurs in a domain wall 
junction~\cite{junction}, \cite{Ito:2000zf}. 
The energy of the intersecting wall receives 
contributions from constituent walls 
and from the junction. 
The known analytic solution in 
Ref.\cite{Ito:2000zf} shows that the 
contribution from the junction is negative. 
This phenomenon may naturally be understood as a kind of 
binding energy of constituent walls. 
}
Since the instanton is stuck at the intersecting point 
of vortices, it may be called an ``intersecton".  
It cannot move once the vortices are fixed. 
We thus conclude that there exist two kinds of instantons; 
one is what lives inside a vortex 
and the other is an instanton 
stuck at the intersection point of vortices. 
As we have seen, there also exists trivially intersecting 
vortices. 
We expect that the most general solution is given by 
the mixture of these configurations.

\medskip
Here we show that the intersectons found above 
essentially exist in $U(1)$ gauge theories.
To this end we consider multiple semilocal vortices 
in the theory with $\NF > \NC$. 
We take the strong gauge coupling 
limit $g^2\rightarrow\infty$ to obtain 
an exact solution. 
In this limit the model
reduces to a nonlinear sigma model 
whose target space is the cotangent bundle over the 
complex Grassmann manifold, $T^* (G_{\NF,\NC})$~\cite{Grassmann}.
Then the master equation (\ref{1/4master}) 
can be solved algebraically as Eq.~(\ref{eq:omega-g-infty}). 
For definiteness we consider a model with $\NC=1$ and $\NF=4$.
The following moduli matrix gives 
non-trivially intersecting vortices with $\nu_{\rm v}=k_z,\ \nu_{\rm v'}=k_w$
\be
 H_0
 = \left(
  z^{k_z}w^{k_w},\ z^{k_z},\ w^{k_w},\ 1
 \right) . \label{U(1)instanton}
\ee
The $\Omega$ can be calculated as 
\be
  \Omega^{g \to \infty}  
  = \Omega_0 = (|z|^2 + 1)^{k_z}(|w|^2 + 1)^{k_w} \; 
\ee
and the exact solution can be obtained as 
$H = (1/ \sqrt{ \Omega^{g \to \infty}} ) H_0$.
The instanton number can be calculated to be the product of 
vorticities, namely $\nu_{\rm i} = - k_z k_w$. 
This solution explicitly shows that 
the $U(1)$ instantons are 
stuck at the intersection of vortices. 
We can show that the instanton charge $\nu_{\rm i}$ 
changes its sign under the duality transformation $\NC$, $\tilde\NC \equiv \NF - \NC$
\cite{Isozumi:2004jc}.
Therefore, there also exist intersectons with
positive instanton charges.

\bigskip
We discuss some more issues in the following. 
Although we ignore $\Sigma$ and zero-th component
of gauge potential $W_0$ in this paper,
we can also construct electrically charged solitons
whose charge is $Q_e=(g^2/2)\int d^4x \ \partial_m(\Sigma F_{0m})$
by restoring these fields. 
The Bogomol'nyi completion gives 
the most general BPS equations 
\be
\D_0\Sigma = 0, \quad 
F_{m0} + \D_m\Sigma = 0,\quad 
\D_0H^i +i\Sigma H^i =0,
\label{dyonic}
\ee
added to Eq.(\ref{eq:1/4bps1}) and (\ref{eq:1/4bps2}).
These equations have to be solved with the Gauss's law
\be
\frac{1}{g^2}\D_mF_{m0} = \frac{i}{g^2}\left[\Sigma,\D_0\Sigma\right]
+ \frac{i}{2}\left[H\D_0H^\dagger - \D_0HH^\dagger\right].
\label{gauss}
\ee
We can show that solutions of 
these equations are $1/4$ BPS states.
If we turn off the FI paramter $c$ and set $H^i=0$, 
these BPS solutions reduces to that for 
the 1/2 BPS dyonic instanton\cite{dyonic}.
The time-independent solutions 
of the dyonic instantons in the Higgs phase
can be obtained as follows.
First we solve the BPS equations (\ref{eq:1/4bps1})
and (\ref{eq:1/4bps2}) as shown in this paper.
Next we set $\partial_0=0$ and $W_0 = -\Sigma$ to solve additional
equations (\ref{dyonic}). Finally $\Sigma$ can be obtained by 
solving the Gauss's law under a given solution of instantons
in the Higgs phase as the background
\be
\D_m\D_m \Sigma = -\frac{g^2}{2}\left(HH^\dagger\Sigma + \Sigma HH^\dagger\right).
\ee
In superstring theory 
dyonic instantons in the pure SUSY Yang-Mills theory
were found to be supertubes \cite{Mateos:2001qs}
(and see also references in \cite{Townsend:2004nc})
between parallel D4-branes \cite{supertube}.
Brane constructions for the dyonic instantons 
in the Higgs phase is an open problem.

\bigskip
Our instantons in the Higgs phase share 
some properties with non-commutative instantons. 
First $U(1)$ instantons can exist in the Higgs phase 
as shown in Eq.~(\ref{U(1)instanton}) 
like non-commutative $U(1)$ instantons \cite{NS}.
Second their topologies are similar (but not identical)
as stated in the footnote \ref{non-commutative}. 
The moduli space of (non-commutative) instantons 
has the hyper-K\"ahler structure.
Although the moduli space of the former 
has the K\"ahler structure at least 
and has real dimensions four multiplied by ${\bf Z}$, 
we do not know if it
has the hyper-K\"ahler structure. 
It may be not the case because 
there remain only two SUSY in the former,
but their moduli space may be obtained as 
a deformation of the moduli space of the latter preserving 
only the K\"ahler structure. 
It is very interesting to explore 
more simiralities between instantons in the Higgs phase 
and non-commutative instantons~\cite{NS}.
It is also desired to obtain the ADHM construction for 
our instantons in the Higgs phase.

\bigskip
Dynamics of instantons within a vortex is equivalent to
the dynamics of lumps~\cite{Ward:1985ij,Stokoe:1986ic}. 
Further studys in dynamics of sigma model lumps 
would clarify dynamics of instantons 
in more general configurations. 
Not only classical dynamics but also
quantum effects in these solitons are important subjects.
It was found by N.~Dorey in \cite{Dorey:1998yh} 
that the BPS spectra in $d=4$ ${\cal N}=2$ 
SUSY gauge theory with $\NF > \NC$ and 
$d=2$ ${\cal N}=(2,2)$ SUSY ${\bf C}P^N$ 
model with twisted masses completely coincide. 
One explanation for this coincidence 
has been given in \cite{Hanany:2004ea,Shifman:2004dr} 
by considering a monopole inside a vortex. 
In conformity with these observations, 
there exist similarities between 
$d=4$ Yang-Mills instantons and 
$d=2$ sigma-model instantons (lumps).
Our results in the present paper 
give a further evidence for the relation because 
we have realized instantons inside a vortex 
as sigma model lumps on the vortex theory.
For instance, small instanton singularities are 
understood as small lump singularities. 
It is also quite interesting to generalize 
the instanton couting~\cite{Nekrasov:2002qd} to 
the case of the instantons in the Higgs phase.

Non-Abelian walls found in \cite{Isozumi:2004jc}
are recently realized as 
D-brane configurations in string theory~\cite{EINOOST}. 
By doing this the diverse phenomena of non-Abelian walls
can be easily understood by dynamics of D-branes.
Monopoles in the Higgs phase are 
also realized by the same brane 
configuration~\cite{Hanany:2004ea,Auzzi:2004yg}. 
Hence we would like to realize the instantons in 
the Higgs phase by some D-brane configuration, 
and we expect that 
the relation between monopoles and instantons in the Higgs phase 
can be interpreted as a T-duality 
in such a D-brane configuration, 
as in the case of the ordinary monopoles and instantons.

\section*{Acknowledgements}
We would like to thank Katsushi Ito and David Tong 
for a useful discussion. 
We also thank the Yukawa Institute for 
Theoretical Physics at Kyoto University. 
Discussions during the YITP workshop YITP-W-04-03 on 
``Quantum Field Theory 2004'' were useful to complete this work. 
This work is supported in part by Grant-in-Aid for Scientific 
Research from the Ministry of Education, Culture, Sports, 
Science and Technology, Japan No.13640269 (NS) 
and 16028203 for the priority area ``origin of mass'' 
(NS). 
The works of K.O. and M.N. are 
supported by Japan Society for the Promotion 
of Science under the Post-doctoral Research Program. 
M.E. and Y.I. gratefully acknowledge 
support from a 21st Century COE Program at 
Tokyo Tech ``Nanometer-Scale Quantum Physics" by the 
Ministry of Education, Culture, Sports, Science 
and Technology.  
M.E. gratefully acknowledges 
support from the Iwanami Fujukai Foundation.


\end{document}